\documentclass[aps,pre,amsmath,lengthcheck,superscriptaddress]{revtex4-2}

\usepackage{graphicx}\graphicspath{ {figures/} }
\usepackage[caption=false]{subfig}
\usepackage[colorlinks=true,allcolors=blue]{hyperref}

\newcommand{\pd}{\partial}
\newcommand{\bra}[1]{\langle #1|}
\newcommand{\ket}[1]{|#1\rangle}
\newcommand{\braket}[2]{\langle #1 | #2 \rangle}

\newcommand{\tr}[1]{\mathrm{Tr}\left\{ #1 \right\}}
\newcommand{\field}{{B}}

\begin{document}

\title{Failure of the geometric approach prediction of excess work scaling for open and isolated quantum systems}

\author{Artur Soriani}
\email{asorianialves@gmail.com}

\author{Eduardo Miranda}

\author{Marcus V. S. Bonan\c{c}a}
\email{mbonanca@ifi.unicamp.br}

\affiliation{Gleb Wataghin Institute of Physics, University of Campinas, Campinas, São Paulo 13083--950, Brazil}

\date{\today}

\begin{abstract}

The task of finding optimal protocols that minimize the energetic cost of thermodynamic processes of long yet finite duration $\tau$ is a pressing one.
We approach this problem here in a rigorous and systematic fashion by means of the adiabatic perturbation theory of closed Hamiltonian quantum systems.
Our main finding is a $1/\tau^2$ scaling of the excess work for large $\tau$ in gapped systems.
This result is at odds with the asymptotic $1/\tau$ prediction of the geometric approach to optimization, which is predicated on the slow evolution of open systems close to canonical equilibrium.
In contrast, our approach does not lead to an obvious geometric interpretation.
Furthermore, as the thermodynamic work does not depend on how an isolated quantum system is split into a system of interest and its environment, our results imply the failure of the geometric approach prediction even for open systems.
Additionally, we provide alternative optimization procedures, both for slowly-varying processes described by adiabatic perturbation theory and for weakly-varying processes described by linear response theory.
Our findings are benchmarked and confirmed through the application to the driven transverse-field Ising chain.

\end{abstract}

\maketitle

\section{Introduction \label{sec:Introduction}}

In the last decades, we have witnessed the development of several experimental techniques that enable the control and manipulation of few atoms, molecules or particles at the nanoscale.
Using optical tweezers \cite{blickle2012natphys,martinez2016natphys,kumar2020nature}, ion traps \cite{stick2005natphys}, nuclear magnetic resonance \cite{vander2005control}, optical lattices \cite{chu2002control} and other techniques \cite{koch2019control,rademacher2022prl}, different experimental setups have been implemented to study several kinds of control and applications with potential development of new technologies.

The need to refine the manipulation of such small systems, increasing the level of control and decreasing the corresponding costs, has motivated intense theoretical activity on the topic.
Among the prominent theoretical proposals are those of shortcuts to adiabaticity \cite{odelin2019rmp,rice2003jpc,berry2009jpa,chen2010prl,chen2011pra,campo2012prl,torrontegui2012pra,jarzynski2013pra,deffner2014prx,masuda2014prl,deffner2016njp}, which encompass many related methods.
They all have the same goal, namely, the implementation of finite-time controls such that the final state of the system of interest is exactly identical to that produced by an equivalent quasistatic manipulation.

The remarkably good results of this theoretical framework do not hide, however, a few limitations.
The implementations often require additional control mechanisms \cite{odelin2019rmp,rice2003jpc,berry2009jpa} or specific characteristics of the control fields \cite{deffner2014prx,deng2018pra}, restricting the possible setups in the laboratory.
Extensions to many-body systems may also face some difficulties, although it has been proved possible in some cases \cite{deng2018pra,saberi2014pra,campbell2015prl,sels2017pnas,Soriani2022PRL}.
The different shortcuts to adiabaticity have been mostly restricted to isolated systems, i.e., systems whose dynamics is Hamiltonian.
However, some extensions to open quantum systems have surfaced \cite{vacanti2014njp,alipour2020quantum,santos2021pra}.
In addition, only recently the issue of quantifying the cost to implement these control techniques has been addressed in the literature \cite{zheng2016pra,campbell2017prl}.

In parallel to this, the research on optimal finite-time thermodynamic processes has been seeking similar goals, although using a different perspective \cite{martinezNP2016,liPRE2017,pancottiPRX2020,liPRL2022,frimPRE2022}.
In this field, the minimization of the energetic cost during the finite-time control is the main target.
In light of what was just presented, we might ask ourselves whether optimal finite-time control is also a kind of approximate shortcut to adiabaticity, and hence whether the two approaches can complement each other \cite{deffner2020epl}.
In the classical regime, the geometric approach to optimal slowly-varying processes \cite{Sivak2012,Bonanca2014} has been successfully applied to several interesting situations \cite{zulkowski2012pre,zulkowski2015pre,rotskoff2016pre,sivak2016pre,rotskoff2017pre,lucero2019pre,blaber2020jcp,louwerse2022preprint,Blaber2022,Abiuso2022,wadiaPRE2022}.
It has been also generalized to the quantum regime when the system of interest is weakly interacting with a heat bath \cite{Zulkowski2015,Scandi2019,abiuso2020prl,abiuso2020entropy,alonso2022prxq}.
In this approach, the energetic cost is expressed as an integral of a Lagrangian that is understood as a metric.
The protocols minimizing such functional are then the corresponding geodesics. 

In the present work, we show that the geometric approach cannot be applied to isolated quantum systems with a gap.
We prove this rigorously through adiabatic perturbation theory (APT) \cite{Rigolin2008}: for a slowly-varying process of duration $\tau$, APT provides a systematic perturbative expansion in powers of $\tau^{-1}$.
The energetic cost is shown to decay as $\tau^{-2}$, in clear contrast to the asymptotic $\tau^{-1}$ prediction of the geometric approach \cite{Sivak2012,Bonanca2014,Zulkowski2015,Scandi2019,abiuso2020entropy}.
Additionally, we show that the latter cannot be applied to open quantum systems either, when the underlying microscopic dynamics of system plus environment is Hamiltonian, provided the full system is gapped.
Our approach also allows for the optimization of slow processes.
Taking the quantum Ising chain at zero temperature as a benchmark test, we compare our theoretical predictions with numerical simulations of the exact time-dependent dynamics, with excellent agreement.
We also provide optimal finite-time protocols, following Ref.~\cite{bonanca2018pre}, in the complementary regime of fast but weak processes, where the variation of the external parameter is small compared to its initial value.

We will describe in detail the conditions in which our results apply.
Foremost among them is the presence of a spectral gap.
At this point, we just point out that gapped many-body systems are very common.
Of course, in the important case of small systems, there are always finite-size gaps.
However, even in the thermodynamic limit, gapped systems abound: (a) conventional $s$-wave BCS superconductors~\cite{Parks1969}, (b) quantum Hall systems (both integer and fractional)~\cite{Girvin1999}, (c) integer-spin (Haldane) chains~\cite{Haldane1983}, and (d) quantum disordered paramagnets~\cite{Sachdev2011}, to name a few.
The quantum Ising model, whose one-dimensional version we study here, is just one example of a wide class of magnetic systems that are gapped \textit{except at a point} in the phase diagram, namely, its critical point [see Ref.~\cite{Sachdev2011} for numerous examples].
In fact, gapless behavior often relies on additional conditions, as (a) the presence of exact symmetries, which give rise to Goldstone modes in phases with a spontaneously broken continuous symmetry or (b) some topological effect, as in half-integer spin chains~\cite{Haldane1983} or gapless quantum spin liquids~\cite{Broholm2020}.
Departures from exact symmetries, as provided, e.g., by spin anisotropies induced by spin-orbit interaction, frequently end up generating a finite gap in these systems.

The performance of the two perturbative approaches applied here is tested far from equilibrium, and a clear qualitative difference is observed in the optimal protocols when the transition from one regime to the other is made.
This highlights that different optimization strategies are required in different regions far from equilibrium.
Only two parameters are necessary to delimit such regions, namely, the relative change of the control parameter $\lambda$ (which is the transverse field in the case of the Ising chain) and the ratio between a characteristic time scale of the system and the duration $\tau$ of the process.

We show that the leading-order expressions for the energetic cost are quadratic forms in the speed $\dot{\lambda}$ in both perturbative approaches \cite{bonanca2018pre,Soriani2022PRA}.
However, the notion of a metric, as it exists in the geometric approach, seems to be absent.
In the regime of fast but weak processes, the reason for that is related to the memory kept along the process by the generalized force conjugate to $\lambda$.
In other words, the average force at a given time $t_0$ depends not only on $\lambda(t_0)$, but on the whole history of the protocol $\lambda(t)$ up to $t_0$.
In the regime of slowly-varying processes, adiabatic perturbation theory provides a very simple quadratic form that depends only on the initial and final values of $\dot{\lambda}$.
It is interesting to note that the metric of the geometric approach and the two quadratic forms we obtain here are derived from the same object, namely, an autocorrelation function of the generalized force.

The presentation is organized as follows.
We briefly review the basics of the geometric approach to finite-time thermodynamics in Sec.~\ref{sec:GeometricApproach}.
The thermodynamic work in isolated systems is presented in Sec.~\ref{sec:ExcessWork}, and in Sec.~\ref{sec:SlowProcesses} we describe in detail the behavior that adiabatic perturbation theory predicts for the excess work in slow processes, highlighting the differences to the geometric approach and showing possible avenues of optimization.
The complementary regime of weak processes is then discussed in Sec.~\ref{sec:WeakProcesses}, together with the recently proposed method of optimization based on linear response theory.
We briefly discuss the application of the presented methods to nonintegrable systems in Sec.~\ref{sec:NonIntegrable} and the failure of the $\tau^{-1}$ scale even for open systems in Sec.~\ref{sec:OpenSystems}. A discussion of the results and the assumptions we made to obtain them is given in Sec.~\ref{sec:Discussion} and the paper is concluded in Sec.~\ref{sec:Conclusion}.

\section{Geometric approach \label{sec:GeometricApproach}}

Consider a system with a collection of external parameters $\boldsymbol\lambda$ to be varied between times $t_i$ and $t_f$, where $\tau = t_f - t_i$ is the process duration.
A specific choice of the time dependence of $\boldsymbol\lambda(t)$ is called a protocol, and here we only consider protocols that can be written as a function of $t/\tau$.
In any process, the average work $W$ delivered to the system can be separated in two parts,
\begin{equation} \label{eq:AverageWork}
W(\tau) = W_\mathrm{qs} + W_\mathrm{ex}(\tau)
\end{equation}
where $W_\mathrm{qs}$ is the $\tau$-independent (and protocol independent) quasistatic contribution, while $W_\mathrm{ex}$ is the excess contribution, which embodies the extra energy we must give to the system in order to the carry out the process in finite time.
The first part, $W_\mathrm{qs}$, only depends on the end points of the process and represents, therefore, an inescapable energetic cost;
the second part, $W_\mathrm{ex}$, is protocol-dependent and can be minimized with a clever choice of \textit{how} we drive the system.
No matter how it is done, however, $W_\mathrm{ex} \to 0$ as $\tau \to \infty$, the adiabatic (quasistatic) limit.

The original formulation of the geometric approach to optimal driving \cite{Sivak2012,Bonanca2014} employs linear response theory (LRT) to describe the system's dissipation for long but finite process duration.
It equates the excess work to
\begin{equation} \label{eq:GeoExcessWork}
W_\mathrm{ex}(\tau) = \int_{t_i}^{t_f} \dot{\boldsymbol\lambda}^T(t) \cdot \boldsymbol\zeta(\boldsymbol\lambda(t)) \cdot \dot{\boldsymbol\lambda}(t) dt,
\end{equation}
where overdots denote time derivatives and the superscript $T$ denotes the transpose of a matrix.
The function $\boldsymbol\zeta(\boldsymbol\lambda)$ is the aptly named friction tensor, which is large for values of $\boldsymbol\lambda$ where the system dissipates more energy.
Although Eq.~\eqref{eq:GeoExcessWork} was initially obtained for classical systems, it should also be valid for quantum systems, as long as $\boldsymbol\zeta$ is suitably defined from LRT's quantum formulation of the response function.
Nevertheless, it is also possible to derive an equation with the same quadratic-in-$\dot{\boldsymbol\lambda}$ form of Eq.~\eqref{eq:GeoExcessWork} starting from a strictly quantum Lindblad master equation \cite{Zulkowski2015,Scandi2019} and arriving at a different definition of $\boldsymbol\zeta$.

The friction has a clear geometric interpretation: it induces a Riemannian metric in parameter space, and the geodesics of such space (calculated with straight application of Euler-Lagrange equations) are paths of least resistance, i.e., protocols that minimize dissipation \cite{Sivak2012}.
Additionally, Eq.~\eqref{eq:GeoExcessWork} predicts that the excess work behaves asymptotically as $\tau^{-1}$, as can be readily seen with a change of integration variable to $s = (t-t_i)/\tau$.
Of course, this is consistent with the vanishing of the excess work in the adiabatic limit.
Nevertheless, it predicts a regime in which the decay of $W_\mathrm{ex}$ is universally given by $\tau^{-1}$ regardless of the shape of $\boldsymbol{\lambda}(t)$.

Both the LRT and the Lindblad derivation of Eq.~\eqref{eq:GeoExcessWork} assume that the system of interest is in contact with a heat bath.
They further assume that the slow evolution never takes the system too far away from canonical equilibrium.
In what follows, we show that, for thermally insulated quantum systems with equilibrium initial states, the description of the excess work for slow processes is remarkably different.
More precisely, we prove that, within the region of validity of our approach, the $\tau^{-1}$ behavior of the excess work for large $\tau$ does not exist in this context.

\section{Excess work for isolated systems \label{sec:ExcessWork}}

Consider an isolated quantum system described by Hamiltonian $H(\lambda)$ and evolving in time under unitary dynamics.
Define the instantaneous eigen-equation $H(\lambda) \ket{n(\lambda)} = E_n(\lambda) \ket{n(\lambda)}$.
For simplicity, we focus on  systems with discrete non-degenerate spectra (this assumption can be relaxed) and restrict ourselves to the case of a single external parameter $\lambda$, to be varied from $\lambda_i$ to $\lambda_f$.
Assuming that the system's initial density matrix is an equilibrium distribution, $\rho(t_i) = \sum_n p_n \ket{n(\lambda_i)} \bra{n(\lambda_i)}$, the density matrix at time $t$ is
\begin{equation} \label{eq:DensityMatrix}
\rho(t) = U(t,t_i) \rho(t_i) U^{\dag}(t,t_i) = \sum_n p_n \ket{\psi_n(t)} \bra{\psi_n(t)},
\end{equation}
where $U$ is the unitary time evolution operator and $\ket{\psi_n(t)} = U(t,t_i) \ket{n(\lambda_i)}$ is the solution to Schrödinger's equation with initial condition $\ket{\psi_n(t_i)} = \ket{n(\lambda_i)}$.
To have an initially canonical distribution, one would choose $p_n = Z_i^{-1} e^{-\beta_i E_n(\lambda_i)}$, where $\beta_i$ is the initial inverse temperature and $Z_i$ is the associated partition function.

Defining the average work as the difference between average energies at the end and at the beginning of the process, $W(\tau) = \tr{\rho(t_f) H(\lambda_f) - \rho(t_i) H(\lambda_i)}$, we can use Eq.~\eqref{eq:DensityMatrix} to show that
\begin{equation} \label{eq:AverageTPMWork}
W(\tau) = \sum_{m,n} p_n p_{m|n}(\tau) \bigl( E_m(\lambda_f) - E_n(\lambda_i) \bigr),
\end{equation}
where
\begin{equation} \label{eq:TransitionProbability}
p_{m|n}(\tau) = \left| \braket{m(\lambda_f)}{\psi_n(t_f)} \right|^2
\end{equation}
is a transition probability (or survival probability if $m=n$), i.e., the probability of finding the system in state $\ket{m(\lambda_f)}$ at $t_f$ given that it was in state $\ket{n(\lambda_i)}$ at $t_i$.
In turn, $E_m(\lambda_f) - E_n(\lambda_i)$ is the energy difference of the transition.
The average work, as written in Eq.~\eqref{eq:AverageTPMWork}, is equal to the average of the two-point measurement work, which is the quantum definition of work in a single realization of the process that obeys the quantum versions of Jarzynski's equality \cite{Talkner2007a} and Crook's fluctuation theorem \cite{Talkner2007b}.
Note, however, that this is only the case because we assumed an initial equilibrium state.
Alternatively, the same definition of average work used above can be put in an integrated power form,
\begin{equation} \label{eq:PowerWork}
W(\tau) = - \int_{t_i}^{t_f} \dot{\lambda}(t) \tr{\rho(t) F(\lambda)} dt,
\end{equation}
where we used the generalized force conjugate to $\lambda$,
\begin{equation} \label{eq:GeneralizedForce}
F(\lambda) = - \frac{\pd H(\lambda)}{\pd\lambda}.
\end{equation}
Equation~\eqref{eq:PowerWork} will be useful in the discussion of weak processes of Sec.~\ref{sec:WeakProcesses}.

In the adiabatic limit, the solution to Schrödinger's equation is given by the adiabatic theorem \cite{Messiah1962book}
\begin{equation} \label{eq:AdiabaticTheorem}
\ket{\psi_n^{(0)}(t)} = e^{i \phi_n(t)} \ket{n(\lambda)},
\end{equation}
where we omit the time dependence of $\lambda$ to evince the parametric evolution.
In Eq.~\eqref{eq:AdiabaticTheorem}, the superscript zero signifies that we are in the adiabatic limit and $\phi_n(t)$ contains the usual geometric and dynamic phases,
\begin{equation} \label{eq:AdiabaticPhases}
\phi_n(t) = \int_{t_i}^t \left( i \bra{n(\lambda(t'))} \frac{d}{dt'} \ket{n(\lambda(t'))} - \frac{E_n(\lambda(t'))}{\hbar} \right) dt'.
\end{equation}
In this case, Eq.~\eqref{eq:TransitionProbability} gives $p_{m|n}^{(0)} = \delta_{mn}$ and Eq.~\eqref{eq:AverageTPMWork} gives the quasistatic work,
\begin{equation} \label{eq:QuasiStaticWork}
W_\mathrm{qs} = \sum_{n} p_n \big( E_n(\lambda_f) - E_n(\lambda_i) \big),
\end{equation}
which is notably process independent.

The density matrix of Eq.~\eqref{eq:DensityMatrix}, for a canonical initial distribution and in the adiabatic limit of Eq.~\eqref{eq:AdiabaticTheorem}, reads
\begin{equation} \label{eq:AdiabaticDensityMatrix}
\rho^{(0)}(\lambda) = \sum_n \frac{e^{-\beta_i E_n(\lambda_i)}}{Z_i} \ket{n(\lambda)} \bra{n(\lambda)},
\end{equation}
which is \emph{not} a canonical distribution at time $t$, since the statistical weights are always evaluated at $t_i$.
Therefore, a generic isolated quantum system in a quasistatic process with an initial canonical distribution is taken away from canonical equilibrium at later times.
This contrasts with the assumption of small deviations from canonical equilibrium taken in the geometric approach.
In specific systems where $E_n(\lambda)/E_n(\lambda_i)$ is independent of $n$ (such as the harmonic oscillator with varying frequency or the ideal gas with varying volume), the density matrix keeps its canonical form with a time-dependent temperature, but this is not true in general.
This serves as a hint that the geometric approach is unsuited for isolated systems.

We know from Eq.~\eqref{eq:AverageWork} that the excess work is obtained by subtracting Eq.~\eqref{eq:QuasiStaticWork} from \eqref{eq:AverageTPMWork}.
Using the identity $p_{n|n} = 1 - \sum_{m \neq n} p_{m|n}$, we arrive at
\begin{equation} \label{eq:ExcessWork}
W_{\mathrm{ex}}(\tau) = \sideset{}{'}\sum_{m,n} p_n p_{m|n}(\tau) E_{mn}(\lambda_f),
\end{equation}
where $E_{mn}(\lambda) = E_m(\lambda) - E_n(\lambda)$ and the prime indicates that the diagonal term $m=n$ is not included.
It is noteworthy that the energy difference appearing in Eq.~\eqref{eq:ExcessWork} is evaluated solely on $\lambda_f$.
This means that the parametric energy variation of the eigen-states is entirely accounted for in the quasistatic work of Eq.~\eqref{eq:QuasiStaticWork}, which leaves only the energy variations from transitions (and not survivals) to be considered in the excess work.
Not only does the expression in Eq.~\eqref{eq:ExcessWork} vanish in the adiabatic limit, but the minimal work principle \cite{Allahverdyan2005} ensures that it is always non-negative for non-degenerate systems with initial density matrices obeying $p_n \geq p_m$ if $E_n(\lambda_i) \leq E_m(\lambda_i)$, such as the canonical distribution or the density matrix of a pure ground state. 

Equation \eqref{eq:ExcessWork} contains the expression for the excess work we will consider in slow processes.
Of course, exact evaluation of this equation requires solving Schrödinger's equation to determine the transition probabilities.
In the next section, we introduce adiabatic perturbation theory and its approximate expressions for the excess work.

\section{Excess work in slow processes \label{sec:SlowProcesses}}

Adiabatic perturbation theory \cite{Rigolin2008} provides perturbative corrections to the adiabatic theorem, valid when the process duration $\tau$ is large but finite.
In practice, it gives
\begin{equation} \label{eq:APT_PerturbativeExpansion}
\ket{\psi_n(t)} = \sum_{r = 0}^{\infty} \ket{\psi_n^{(r)}(t)},
\end{equation}
where
\begin{equation} \label{eq:APT_EigenStateExpansion}
\ket{\psi_n^{(r)}(t)} = e^{i\phi_n(t)} \sum_m C_{mn}^{(r)}(t) \ket{m(\lambda)}
\end{equation}
is the $r$th order correction written in the instantaneous basis of $H(\lambda)$.
The expansions of Eqs.~\eqref{eq:APT_PerturbativeExpansion} and \eqref{eq:APT_EigenStateExpansion} are purposefully constructed to include the adiabatic limit in its $r=0$ term, with $C_{mn}^{(0)}(t) = \delta_{mn}$.
As such, Eq.~\eqref{eq:APT_EigenStateExpansion} with $r=0$ is identical to Eq.~\eqref{eq:AdiabaticTheorem}, the adiabatic approximation.
For $r>0$, the expressions of $C_{mn}^{(r)}(t)$ can be systematically calculated \cite{Rigolin2008}.
For example, the expression for $r=1$ and $m \neq n$ is
\begin{equation} \label{eq:APT_FirstOrderCoefficient}
C_{mn}^{(1)}(t) = i\hbar \left( \frac{M_{mn}(t)}{E_{mn}(\lambda)} - e^{i \phi_{mn}(t)} \frac{M_{mn}(t_i)}{E_{mn}(\lambda_i)} \right),
\end{equation}
where $\phi_{mn}(t) = \phi_m(t) - \phi_n(t)$,
\begin{align} \label{eq:APT_Matrix}
M_{mn}(t) = \bra{m(\lambda)} \frac{d}{dt} \ket{n(\lambda)} = \dot{\lambda}(t) \frac{ F_{mn}(\lambda) }{E_{mn}(\lambda)}
\end{align}
and $F_{mn}(\lambda) = \bra{m(\lambda)} F(\lambda) \ket{n(\lambda)}$ are the instantaneous matrix elements of the force, defined in Eq~\eqref{eq:GeneralizedForce}.

Since $\lambda$ is a function of $t/\tau$, it gives $\dot{\lambda} \propto \tau^{-1}$, which sets the order of $C_{mn}^{(1)}(t)$ in Eq.~\eqref{eq:APT_FirstOrderCoefficient}.
Similarly, $C_{mn}^{(2)}(t)$ contains $\ddot{\lambda}$ and $\dot{\lambda}^2$, both of which are proportional to $\tau^{-2}$, and this continues on for higher orders \cite{Rigolin2008}.
Thus, a superscript $(r)$ means that the given quantity is proportional to $\tau^{-r}$.
However, $\tau^{-1}$, by itself, \emph{is not} the proper quantity to determine the validity of the theory.
To assess how accurate first-order APT is, it is more appropriate to consider the inequality
\begin{equation} \label{eq:APT_Condition}
\left\vert C_{mn}^{(1)}(t) \right\vert \ll 1,
\end{equation}
which, looking at Eq.~\eqref{eq:APT_FirstOrderCoefficient}, is always true as long as
\begin{equation} \label{eq:QuantitativeAdiabaticCondition}
\hbar \left\vert \frac{M_{mn}(t)}{E_{mn}(\lambda)} \right\vert \ll 1,
\end{equation}
an inequality known as quantitative adiabatic condition \cite{Tong2005,Tong2010}, a validity condition for the adiabatic theorem itself.

At this point, we describe in more detail the conditions under which we expect our approach to be valid.
From Eq.~(\ref{eq:QuantitativeAdiabaticCondition}), it is clear that APT breaks down if some relevant $E_{mn}$ vanishes.
However, there are some important cases in which this will \textit{not} happen.
First, in finite systems, the energy differences $E_{mn}$ are bounded from below by finite-size gaps.
Second, systems that are gapped in the thermodynamic limit (see several examples in Sec.~\ref{sec:Introduction}) will not violate Eq.~(\ref{eq:QuantitativeAdiabaticCondition}) if they are driven at $T=0$ or at temperatures much smaller than the gap.
This can be seen from Eq.~(\ref{eq:ExcessWork}), where it is clear that transitions other than from the ground state are absent or strongly suppressed by the Boltzmann weight.
This will be the case of all our numerical simulations later on.

In possession of Eq.~\eqref{eq:APT_PerturbativeExpansion}, one can easily write down the transition probabilities appearing in Eq.~\eqref{eq:ExcessWork}.
We have
\begin{equation} \label{eq:APT_TransitionProbability}
p_{m|n}(\tau) = \left| \sum_{r=0}^{\infty} C_{mn}^{(r)}(t_f) \right|^2
\end{equation}
and, in practice, one calculates Eq.~\eqref{eq:APT_TransitionProbability} to the desired order.
We are now able to determine the corrections to the excess work coming from the corrections to the time-dependent state.
It must be pointed out that, since the sum in Eq.~\eqref{eq:APT_TransitionProbability} is squared, a specific correction $C_{mn}^{(r)}$ may emerge in many orders of correction to $W_{\mathrm{ex}}$, not only in order $r$.

We already know that the $r=0$ term of the theory reproduces the quasistatic work, and thus does not contribute to the excess work.
Following our superscript convention, truncating Eq.~\eqref{eq:APT_TransitionProbability} at $r=1$ leads to
\begin{equation} \label{eq:APT_W1}
W^{(1)}(\tau) = 0.
\end{equation}
This is the central result of this paper.
According to it, in a slow but finite time process of an isolated gapped system, there is \emph{no} first order correction to the average work in the inverse process duration $\tau^{-1}$.
This result, which relies on the condition of initially diagonal density matrices, contrasts with the geometric approach prediction of  $W_\mathrm{ex} \sim \tau^{-1}$ for slow processes. 
We stress that an initial diagonal state describes an initial thermodynamic equilibrium between the internal parts of the isolated system.
Furthermore, Eq.~\eqref{eq:APT_W1} relies neither on a specific number of degrees of freedom nor on a certain partition of the isolated system.
Thus, it also applies to open quantum systems whose underlying microscopic dynamics with its environment is Hamiltonian and gapped, as further explained in Sec.~\ref{sec:OpenSystems}.

Continuing on for higher orders, we see that the $r=1$ term of Eq.~\eqref{eq:APT_TransitionProbability} is enough to describe the second order correction to the work,
\begin{equation} \label{eq:APT_W2}
W^{(2)}(\tau) = \sideset{}{'}\sum_{m,n} p_n \left| C_{mn}^{(1)}(t_f) \right|^2 E_{mn}(\lambda_f),
\end{equation}
which is the first non-zero correction. Such correction has already been considered in the context of trade-offs between power and efficiency of quantum heat engines~\cite{Chen2019a,Chen2019b}. Equation \eqref{eq:APT_W2} is the starting point for the optimization of the energetic cost of slow processes in isolated quantum systems.

\subsection{Optimizing the excess work}

In contrast to Eq.~\eqref{eq:GeoExcessWork}, APT results for the excess work have no functional dependence: Eq.~\eqref{eq:APT_W2} only depends on $\lambda$ and $\dot{\lambda}$ at $t_i$ and $t_f$, not on the entire history of the protocol $\lambda(t)$.
In fact, if $\dot{\lambda}(t_i) = 0 = \dot{\lambda}(t_f)$, Eq.~\eqref{eq:APT_W2} also vanishes (see Eqs.~\eqref{eq:APT_FirstOrderCoefficient} and \eqref{eq:APT_Matrix}) and the first non-zero correction would be
\begin{equation} \label{eq:APT_W4}
W^{(4)}(\tau) = \sideset{}{'}\sum_{m,n} p_n \left| C_{mn}^{(2)}(t_f) \right|^2 E_{mn}(\lambda_f),
\end{equation}
where $C_{mn}^{(2)}(t)$ is similar to $C_{mn}^{(1)}(t)$ of Eq.~\eqref{eq:APT_FirstOrderCoefficient}, differing by the substitution $M_{mn}(t) \to i \hbar \frac{d}{dt} \frac{M_{mn}(t)}{E_{mn}(\lambda)}$.
This continues on for higher orders: zeroing out the first $r$ derivatives of $\lambda$ at the beginning and at the end of the process makes the first non-zero correction to the work be $W^{(2r+2)}$.
This approach to reducing the order of the excess work is known as boundary cancellation method (BCM) \cite{Garrido1962,Sancho1966,Morita2007,Morita2008,Rezakhani2010}.
There is a simple formula for generating a BCM protocol of order $r$, namely \cite{Rezakhani2010}
\begin{equation} \label{eq:BCMprotocols}
\lambda_{\mathrm{BC}r}(t) = \lambda_i + [\lambda_f - \lambda_i] \frac{ \int_{t_i}^{t}  (t_f-t')^r (t'-t_i)^r dt' }{ \int_{t_i}^{t_f} (t_f-t')^r (t'-t_i)^r dt' },
\end{equation}
which is a polynomial for any integer $r > 0$ with all time derivatives up to order $r$ vanishing at $t_i$ and $t_f$.
However, there is a caveat: BCM presupposes the validity of APT.
First, adiabaticity must be secured, and forcing time derivatives of the protocol to be zero at the end points might even be detrimental to reaching that goal.
This is because the protocol will have to ``speed up'' during the process, which might spoil inequalities \eqref{eq:APT_Condition} and \eqref{eq:QuantitativeAdiabaticCondition}.

Conversely, we can abuse inequalities \eqref{eq:APT_Condition} and \eqref{eq:QuantitativeAdiabaticCondition} to find protocols that adhere to APT as soon as possible.
The fast quasi-adiabatic (FQA) strategy \cite{Kastberg1995,Torrontegui2012,Bowler2012,Martinez2013,Martinez2015} amounts to setting the LHS of inequality \eqref{eq:QuantitativeAdiabaticCondition} equal to an initially undetermined constant,
\begin{equation} \label{eq:FQADifferentialEquation}
\hbar \left\vert \frac{M_{mn}(t)}{E_{mn}(\lambda)} \right\vert = c,
\end{equation}
and solving this first-order differential equation for $\lambda$ with boundary conditions $\lambda(t_i) = \lambda_i$ and $\lambda(t_f) = \lambda_f$.
Since there is only one constant of integration, $c$ must also be used to impose the boundary conditions, which ultimately leads to $c \propto \tau^{-1}$.
Comparison of Eq.~\eqref{eq:FQADifferentialEquation} to Eq.~\eqref{eq:APT_FirstOrderCoefficient} leads to the conclusion that the FQA protocol distributes first-order transition probabilities uniformly throughout the entire duration of the process.

As it is, Eq.~\eqref{eq:FQADifferentialEquation} only works when considering a single pair of energy levels $m$ and $n$, but we can use the expressions for the work in Eqs.~\eqref{eq:QuasiStaticWork} and \eqref{eq:APT_W2} to take every level pair into account.
To this end, and inspired by Eq.~\eqref{eq:FQADifferentialEquation}, we set the ratio $| W^{(2)}/W^{(0)} |$ equal to a to-be-determined constant $c$ and disregard the terms that are not fully $t$-dependent, which leads to
\begin{equation} \label{eq:FQAAltDifferentialEquation}
\hbar^2 \left\vert \sideset{}{'}\sum_{m,n} p_n \frac{\left| M_{mn}(t) \right|^2}{E_{mn}(\lambda)} \right\vert = c \left\vert \sum_n p_n E_n(\lambda) \right\vert.
\end{equation}
Equation \eqref{eq:FQAAltDifferentialEquation} should then be solved with $\lambda(t_i) = \lambda_i$ and $\lambda(t_f) = \lambda_f$, a strategy which we also refer to as FQA for consistency.
In contrast to BCM, FQA does not cancel Eq.~\eqref{eq:APT_W2}, but it may guarantee the validity of such equation for the smallest $\tau$ possible.

In essence, BCM is an answer to the problem ``given APT validity, find protocols that minimize the cost order by order''.
Conversely, FQA is an answer to the problem ``find protocols that ensure APT validity and are as fast possible'', ``fast'' meaning small process duration.
Both strategies have their advantages, as we shall see next.
In the following analysis, it will be useful to write the external parameter as
\begin{equation} \label{eq:LambdaIntoG}
\lambda(t) = \lambda_i + \Delta\, g(t),
\end{equation}
where $\Delta \equiv \lambda_f - \lambda_i$ is the total variation of $\lambda$ in the process and $g(t)$ is a function obeying $g(t_i) = 0$ and $g(t_f) = 1$.

\begin{figure*}

\subfloat[\label{fig:APT_para_ptc}]{\includegraphics[width=.33\textwidth]{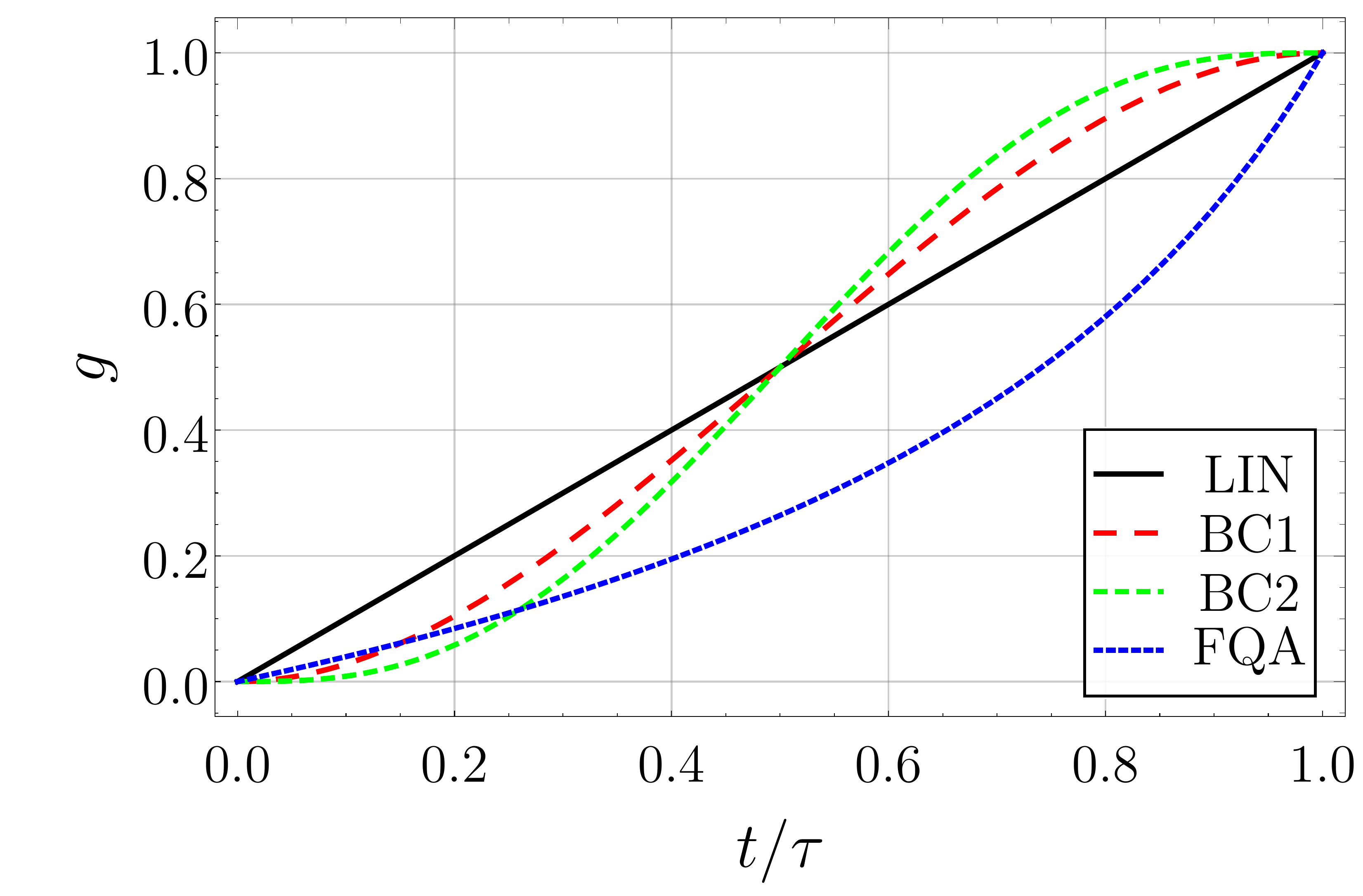}}
\subfloat[\label{fig:APT_para_work_tau}]{\includegraphics[width=.33\textwidth]{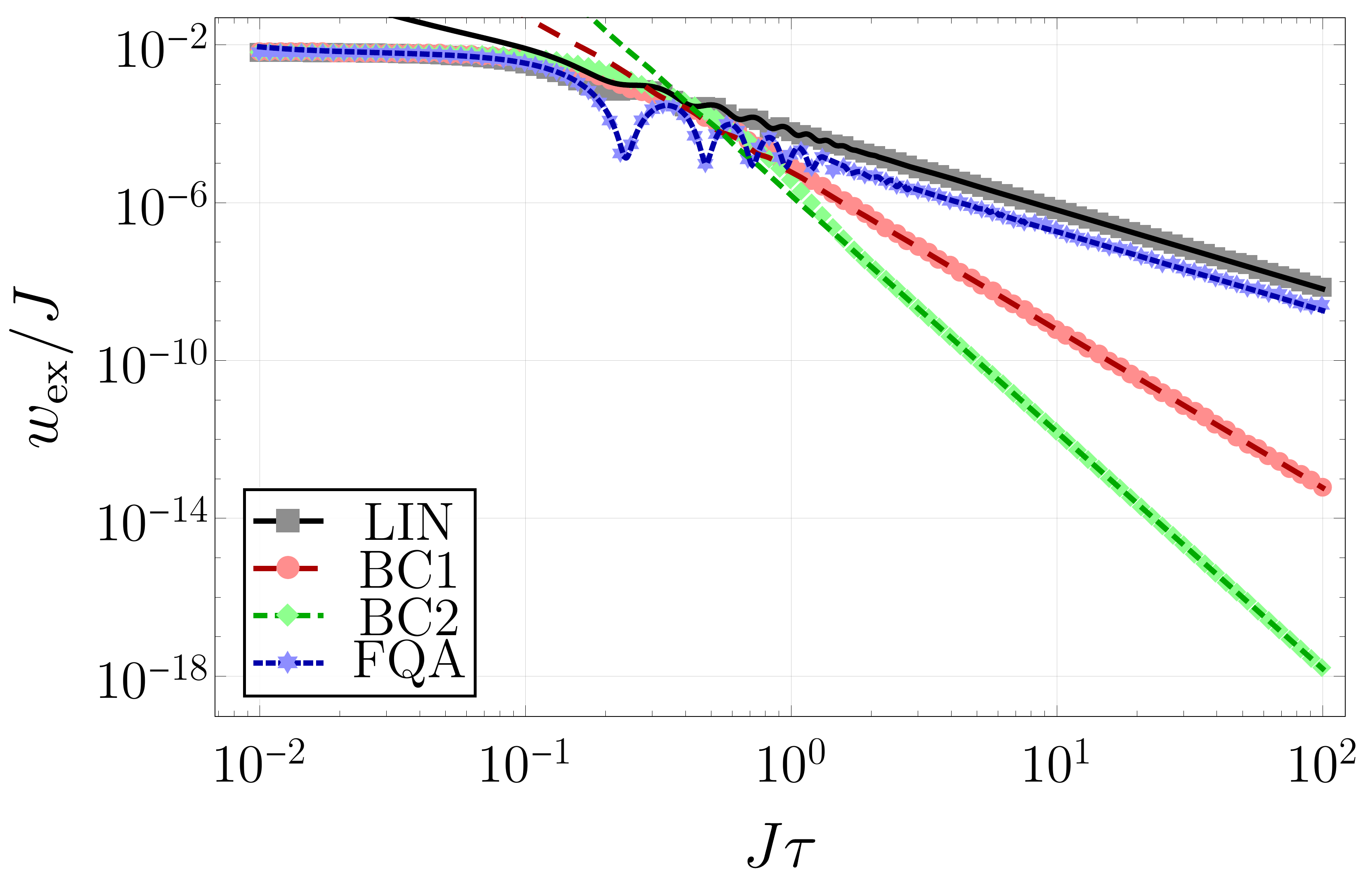}}
\subfloat[\label{fig:APT_para_work_Delta}]{\includegraphics[width=.33\textwidth]{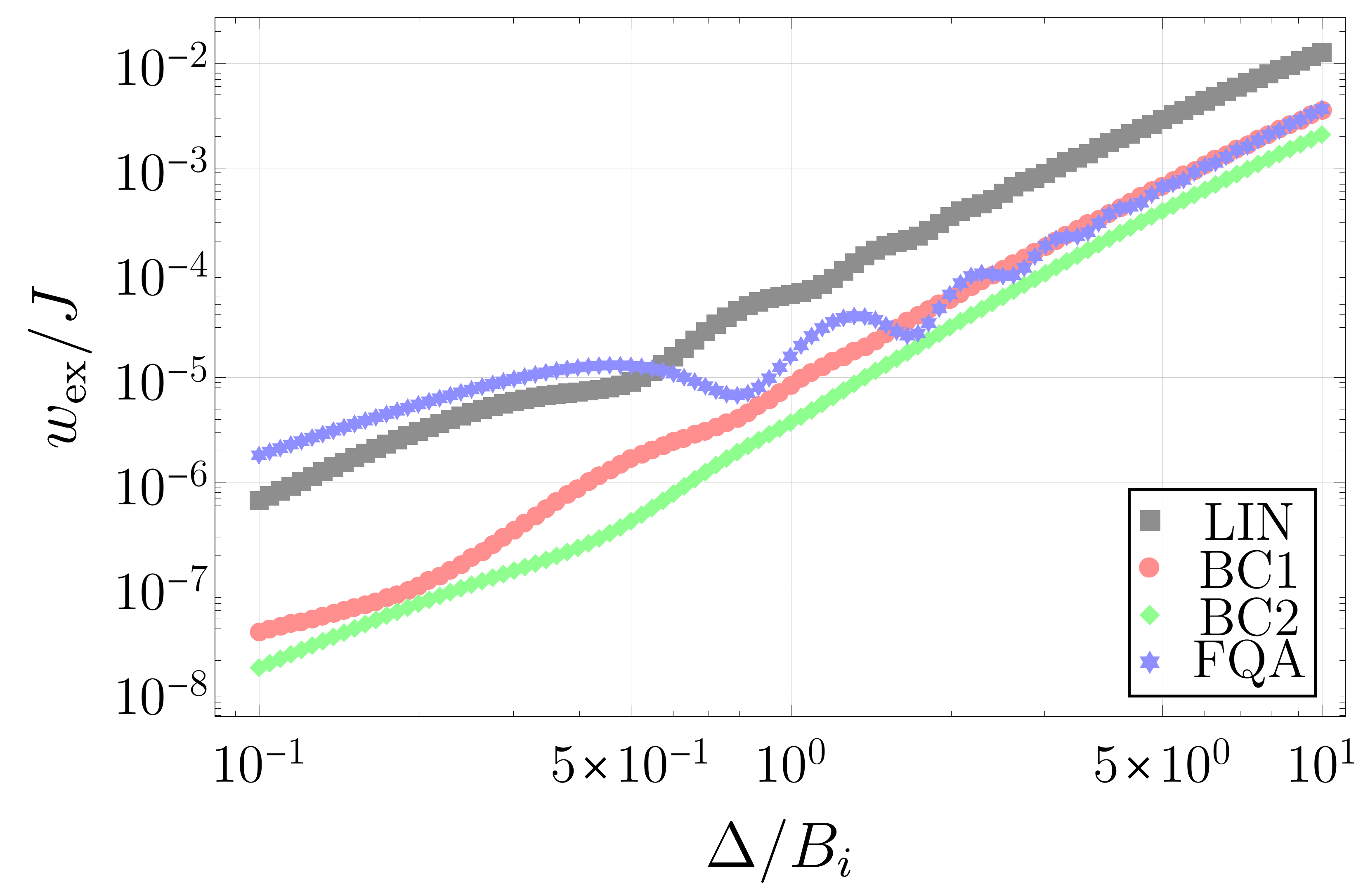}}

\subfloat[\label{fig:APT_ferro_ptc}]{\includegraphics[width=.33\textwidth]{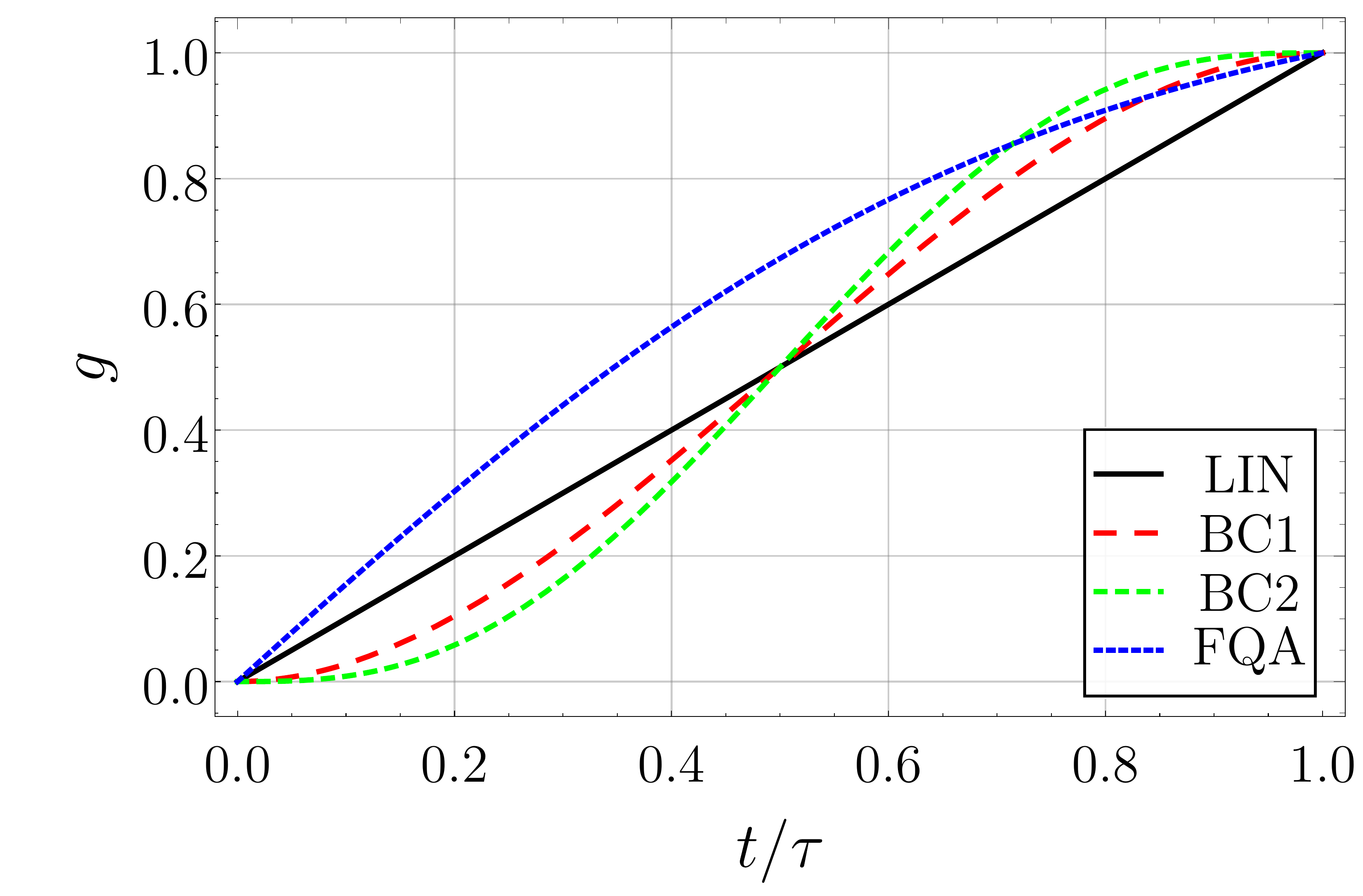}}
\subfloat[\label{fig:APT_ferro_work_tau}]{\includegraphics[width=.33\textwidth]{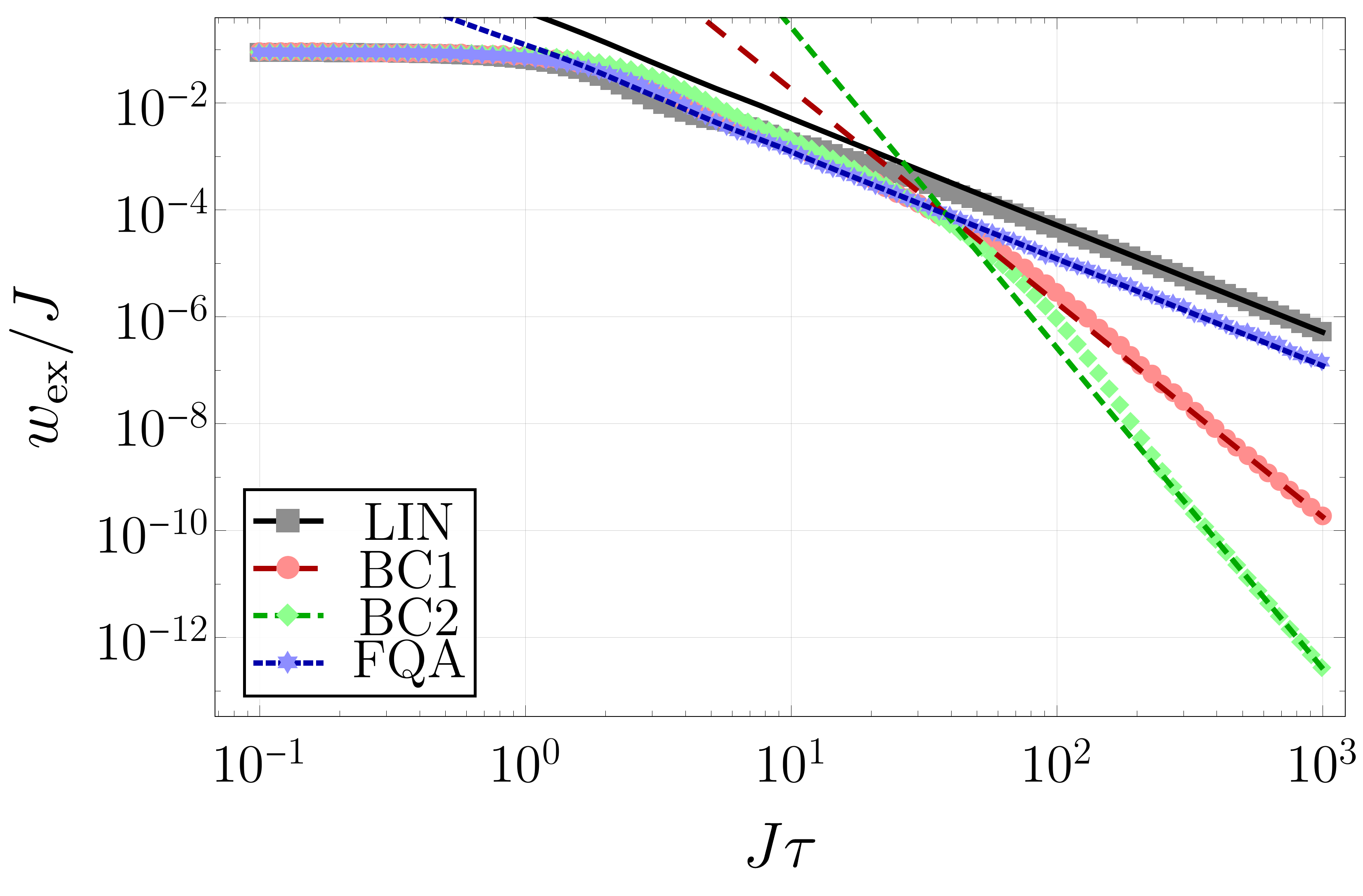}}
\subfloat[\label{fig:APT_ferro_work_Delta}]{\includegraphics[width=.33\textwidth]{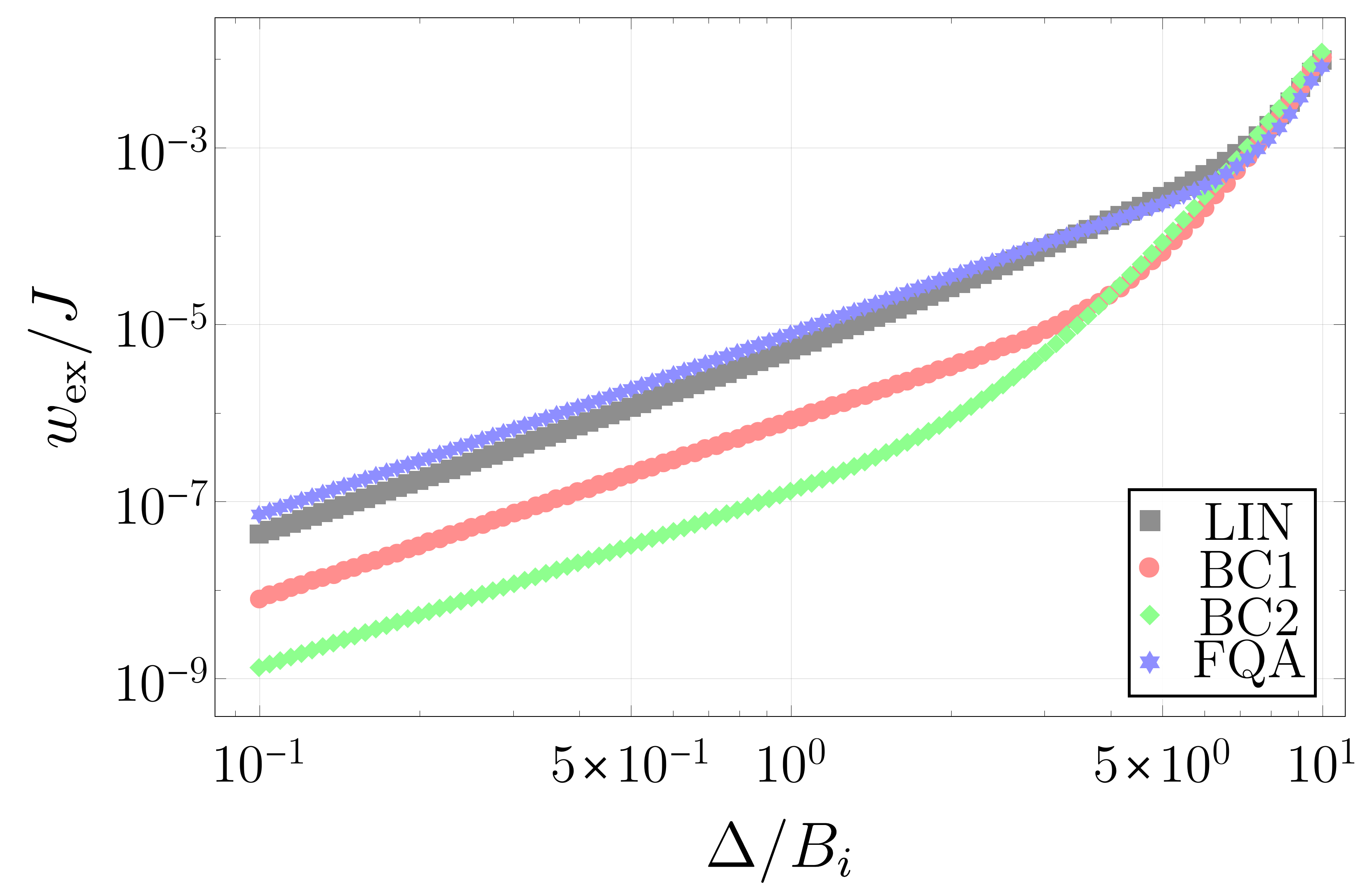}}

\caption{\label{fig:APT_protocols}
Driving protocols and associated excess work per spin $w_\mathrm{ex} = W_\mathrm{ex}/N$ of Eq.~\eqref{eq:TIwork} for the TI chain with $N=100$ spins.
The symbols in (b), (c), (e) and (f) depict numerical results obtained from the integration of the exact time-dependent dynamics, while the lines in (b) and (e) represent the lowest-order predictions from APT.
The top plots correspond to a paramagnetic process with $\field_i = 10J$, while the bottom plots correspond to a ferromagnetic process with $\field_i = 0.1J$.
LIN corresponds to the linear protocol;
BC1 and BC2 correspond to Eq.~\eqref{eq:BCMprotocols} with $r=1$ and $2$, respectively;
FQA corresponds to the solution of Eq.~\eqref{eq:FQAAltDifferentialEquation}.
(a) Protocols for $\Delta = 10J$;
(b) Excess work vs $\tau$ for $\Delta = 10J$;
(c) Excess work vs $\Delta$ for $J\tau = 1$;
(d) Protocols for $\Delta = 0.8J$;
(e) Excess work vs $\tau$ for $\Delta = 0.8J$;
(f) Excess work vs $\Delta$ for $J\tau = 10$.
}

\end{figure*}

The discussion has been general so far, and any discrete and non-degenerate quantum system can be subjected to any of these strategies.
However, to compare their efficiency, we apply them to a specific model: the transverse field Ising model (TI) \cite{Pfeuty1970}, a one-dimensional chain of $N$ spins, initially prepared in its ground state.
The specifics of the system are detailed in appendix~\ref{sec:TIchain}.
The external parameter is the magnetic field $\lambda = \field$, to be compared with the fixed coupling $J$ between neighboring spins.
Here, for the sake of illustrating our results using this model as an exactly diagonalized testbed with a well defined thermodynamic limit, we consider only processes that do not cross the system's quantum critical point (at $\field = J$), because it brings unwanted difficulties (that we have considered elsewhere \cite{Soriani2022PRL,Soriani2022PRA,Soriani2022nonlinear}).
The results for processes entirely contained in the paramagnetic phase ($\field > J$) or in the ferromagnetic phase ($\field < J$) can be seen in Fig.~\ref{fig:APT_protocols}.

Figure~\ref{fig:APT_para_ptc} shows four driving protocols for a paramagnetic process with $\field_i = 10J$.
LIN is the naive linear protocol, which we will always use as a basic protocol we want to outperform.
BC1 and BC2 are the two lowest order BCM polynomials, given in Eq.~\eqref{eq:BCMprotocols} with $r=1$ and $r=2$, respectively ($r=0$ reproduces LIN).
Lastly, FQA is the solution to Eq.~\eqref{eq:FQAAltDifferentialEquation} and, unlike the other three protocols, it depends on $\Delta$ --- Fig.~\ref{fig:APT_para_ptc} was generated with $\Delta = 10J$.
Figures~\ref{fig:APT_para_work_tau} and \ref{fig:APT_para_work_Delta} show the comparison between numerically obtained exact results (symbols) and APT (lines) predictions for the excess work per spin.

Figure~\ref{fig:APT_para_work_tau} shows the excess work per spin vs $\tau$ and was constructed using the protocols of the previous plot with $\Delta=10J$.
From this figure, it is apparent that the second-order APT result of Eq.~\eqref{eq:APT_W2} for the FQA protocol approximates the dynamics ``early'': already at $J\tau = 10^{-2}$ we see good agreement with the exact result, which is a consequence of FQA's design.
It, however, has the standard $\tau^{-2}$ decay for large $\tau$, since it does not necessarily slow down near $t = t_i$ and $t_f$.
The LIN protocol has the same decay and, in this case, its excess work is always greater than FQA's excess work.
Conversely, among the considered protocols, BC1 and BC2 take the longest to be well approximated by APT, but, once they enter APT regime, they behave as $\tau^{-4}$ and $\tau^{-6}$, respectively.

Figure~\ref{fig:APT_para_work_Delta} complements the previous plot by showing the excess work per spin vs $\Delta$ for an intermediate process duration of $J\tau=1$.
To be clear, while Eq.~\eqref{eq:FQAAltDifferentialEquation} has a different solution for each $\Delta$, the FQA protocol used to generate Fig.~\ref{fig:APT_para_work_Delta} is exactly the one shown in Fig.~\ref{fig:APT_para_ptc}, the same for every value of $\Delta$ plotted --- this was done for ease of representation.
This plot shows how the FQA protocol is sensitive to the energy spectrum: it gives higher excess work than LIN when the field variation is small.
On the other hand, BC2 already outperforms every other protocol shown for this not so large value of $\tau$.

Figures~\ref{fig:APT_ferro_ptc}--\ref{fig:APT_ferro_work_Delta} mirror the analysis of the previous plots, but for a ferromagnetic process with $\field_i = 0.1J$.
Many of the features of the paramagnetic process are repeated in the ferromagnetic process.
Worthy of note is the larger time decades considered in Fig.~\ref{fig:APT_ferro_work_tau} (when compared to Fig.~\ref{fig:APT_para_work_tau}) and an apparent coalescence in the excess work of Fig.~\ref{fig:APT_ferro_work_Delta} for every protocol considered, both of which are consequences of the proximity to the critical point.
Indeed, on the far right of Fig.~\ref{fig:APT_ferro_work_Delta}, the last points in the plot represent processes in which the critical point is crossed, but this is of no consequence to our analysis since this region of the plot is far from the region of validity of the perturbation theory used.

Overall, we can see that the $\tau^{-1}$ behavior of the excess work is never observed as the leading order behavior for slow processes in Fig.~\ref{fig:APT_protocols}.

\section{Excess work in weak processes \label{sec:WeakProcesses}}

To complement the analysis of slowly-varying processes, in this section we study weak processes.
To be specific, ``weak'' means that the total variation $\Delta$ of Eq.~\eqref{eq:LambdaIntoG} is small when compared to initial value of the external parameter, $\lambda_i$.
Such processes are well described by LRT \cite{Kubo1985}, in which the Hamiltonian of the system is expanded to first order in $\Delta$,
\begin{equation} \label{eq:WeakHamiltonian}
H(\lambda) \approx H^{i} - \Delta\,  F^{i} g(t),
\end{equation}
where $H^{i} = H(\lambda_i)$ is the initial Hamiltonian and $F^{i} = F(\lambda_i)$ is the initial force of Eq.~\eqref{eq:GeneralizedForce}.

Equation~\eqref{eq:WeakHamiltonian} is plainly written in standard time-dependent perturbation theory form, whose first order result for the density matrix of Eq.\eqref{eq:DensityMatrix} is \cite{Messiah1962book}
\begin{multline} \label{eq:WeakDensityMatrix}
\rho(t) \approx \rho(t_i) - \frac{\Delta}{i\hbar} \sideset{}{'}\sum_{m,n} [p_n - p_m] F^{i}_{mn} \ket{m_i} \bra{n_i} \\
\times \int_{t_i}^{t} g(t') e^{E_{mn}^{i}[t-t']/i\hbar} dt',
\end{multline}
where $\ket{n_i}$ and $E_{n}^{i}$ are initial eigen-states and eigen-energies.
Then, the average force, itself expanded up to first order as $F(\lambda) \approx F^{i} - \Delta g \pd_\lambda^2 H^{i}$, is
\begin{multline} \label{eq:WeakAverageForce}
\tr{ \rho(t) F(\lambda) } \approx \tr{ \rho(t_i) F^{i}} - \Delta g(t) \tr{ \rho(t_i) \pd_\lambda^2 H^{i}} \\
+ \Delta \int_{t_i}^t g(t) \Phi_i(t-t') dt',
\end{multline}
where we defined the response function
\begin{equation} \label{eq:ResponseFunction}
\Phi(\lambda;t) = \frac{2}{\hbar} \sideset{}{'}\sum_{m,n} p_n \left\vert F_{mn}(\lambda) \right\vert^2 \sin\left( \frac{E_{mn}(\lambda)}{\hbar} t \right)
\end{equation}
and $\Phi_i(t) = \Phi(\lambda_i;t)$.

Finally, the average work is obtained by placing Eq.~\eqref{eq:WeakAverageForce} into the integrated power expression, Eq.~\eqref{eq:PowerWork}.
After an integration by parts of the last term, the boundary term and the previous two terms combined reproduce the weak limit of the quasistatic work in Eq.~\eqref{eq:QuasiStaticWork}.
Thus, the leftover term must be the lowest order contribution to the excess work, which we write as \cite{bonanca2018pre,acconcia2015pre}
\begin{equation} \label{eq:LRTExcessWork}
W_\mathrm{ex}^\mathrm{LRT}(\tau) = \Delta^2 \int_{t_i}^{t_f} \int_{t_i}^{t} \dot g(t) \dot g(t') \Psi_i(t-t') dt' dt.
\end{equation}
The relaxation function $\Psi$ relates to the response function as $\Phi(\lambda;t) = - \pd_t \Psi(\lambda;t)$ and, using Eq.~\eqref{eq:ResponseFunction}, we can write
\begin{equation} \label{eq:RelaxationFunction}
\Psi(\lambda;t) = 2 \sideset{}{'}\sum_{m,n} p_n \frac{ \left\vert F_{mn}(\lambda) \right\vert^2 }{E_{mn}(\lambda)} \cos\left( \frac{E_{mn}(\lambda)}{\hbar} t \right)
\end{equation}
and $\Psi_i(t) = \Psi(\lambda_i;t)$.
It is noteworthy that Eqs.~\eqref{eq:ResponseFunction} and \eqref{eq:RelaxationFunction} display the LRT functions $\Phi$ and $\Psi$ in alternate forms, compared to the usual correlation functions (see Sec.~\ref{sec:NonIntegrable}).
For isolated quantum systems, both forms are equivalent.
Equation~\eqref{eq:LRTExcessWork} is the staring point for the optimization of the energetic cost of weak processes in isolated quantum systems.

\subsection{Optimizing the excess work}

An effective strategy to minimize \eqref{eq:LRTExcessWork} consists of expanding $\dot{g}(t)$ in some basis of functions \cite{bonanca2018pre}.
To do that, we first rewrite Eq.~\eqref{eq:LRTExcessWork} in a more symmetric form,
\begin{equation} \label{eq:LRTExcessWork2}
W_\mathrm{ex}^\mathrm{LRT}(\tau) = \frac{\Delta^2}{2} \int_{t_i}^{t_f} \int_{t_i}^{t_f} \dot g(t) \dot g(t') \Psi_i(t-t') dt' dt.
\end{equation}
where the property $\Psi_{i}(-t)=\Psi_{i}(t)$ was used.
Note the similarities and differences between Eq.~\eqref{eq:LRTExcessWork2} and Eq.~\eqref{eq:GeoExcessWork} of the geometric approach --- both of them present quadratic functional dependence on the first derivative of the external parameter, but the kernel of Eq.~\eqref{eq:LRTExcessWork2}, $\Psi_i(t-t')$, shows explicit time dependence.

The derivatives $\dot{g}(t)$ and $\dot{g}(t')$ can now be properly expanded in the interval $[t_i,t_f]$.
Due to their convenient mathematical properties, the Chebyshev polynomials $T_{l}(x)$ are a good choice of efficient basis of functions \cite{weisse2006kernel}.
Following Refs.~\cite{bonanca2018pre,weisse2006kernel}, the truncated and regularized expansion of $\dot{g}(t)$ in a finite number $n$ of polynomials $T_{l}(x)$ in the interval $[t_i,t_f]$ reads 
\begin{equation}
    \dot{g}(t) = \sum_{l=1}^{n} a_{l}\,g_{n,l}\,T_{l}\left( \frac{ 2t - (t_i + t_f) }{\tau} \right),
    \label{eq:protocol_expansion}
\end{equation}
where $a_{l}$ are the coefficients to be determined in the optimization and the factors $g_{n,l}$ regularize the truncated expansion \cite{weisse2006kernel} to avoid the Gibbs phenomenon at the extremities of the expansion interval. Their expression is \cite{weisse2006kernel}
\begin{multline}
    g_{n,l} = \frac{n-l+1}{n+1}\cos{\left(\frac{\pi l}{n+1}\right)} \\+\frac{1}{n+1} \sin{\left(\frac{\pi l}{n+1} \right)} \cot \left( \frac{\pi}{n+1}\right) .
\end{multline}

Substituting expression \eqref{eq:protocol_expansion} into Eq.~\eqref{eq:LRTExcessWork2}, we obtain
\begin{equation}
    W_\mathrm{ex}^\mathrm{LRT}(\tau) (\Delta^2 \Psi_{i}(0)/2)^{-1} = \sum_{l,j}^{n} A_{lj}a_{l}a_{j}\,,
    \label{eq:LRTExcessWorkExp}
\end{equation}
where the $A_{lj}$ are given by
\begin{multline}
 A_{lj} = g_{n,l} g_{n,j} \int_{t_i}^{t_f} \int_{t_i}^{t_f} \tilde{\Psi}_{i}\left( t-t' \right) \\
 \times T_{l}\left( \frac{ 2t - (t_i + t_f) }{\tau} \right) T_{j}\left( \frac{ 2t' - (t_i + t_f) }{\tau} \right)\, dt' dt,
\end{multline}
with $\tilde{\Psi}_{i}(t) = \Psi_{i}(t)/\Psi_{i}(0)$.

The excess work \eqref{eq:LRTExcessWorkExp} becomes then a finite multidimensional quadratic form in the $a_{n}$ whose minimum we want to find.
The boundary conditions $g(t_i) = 0$ and $g(t_f) = 1$ are additional constraints in our optimization problem.
Using the method of Lagrange multipliers, we can find then the coefficients $a_{l}$ that provide the optimal protocol by solving a set of linear algebraic equations.
Naturally, we call this optimization strategy Chebyshev expansion up to $n$ modes (CE$n$).
We remark that the relaxation function $\Psi_{i}(t)$ is the main physical input in this procedure.
The factors $A_{lj}$ crucially depend on the switching time $\tau$ and $\Psi_{i}(t)$.
Due to its relation with the response function, $\Psi_{i}(t)$ can be obtained from experiments when it is not accessible theoretically.

\begin{figure}

\includegraphics[width=\columnwidth]{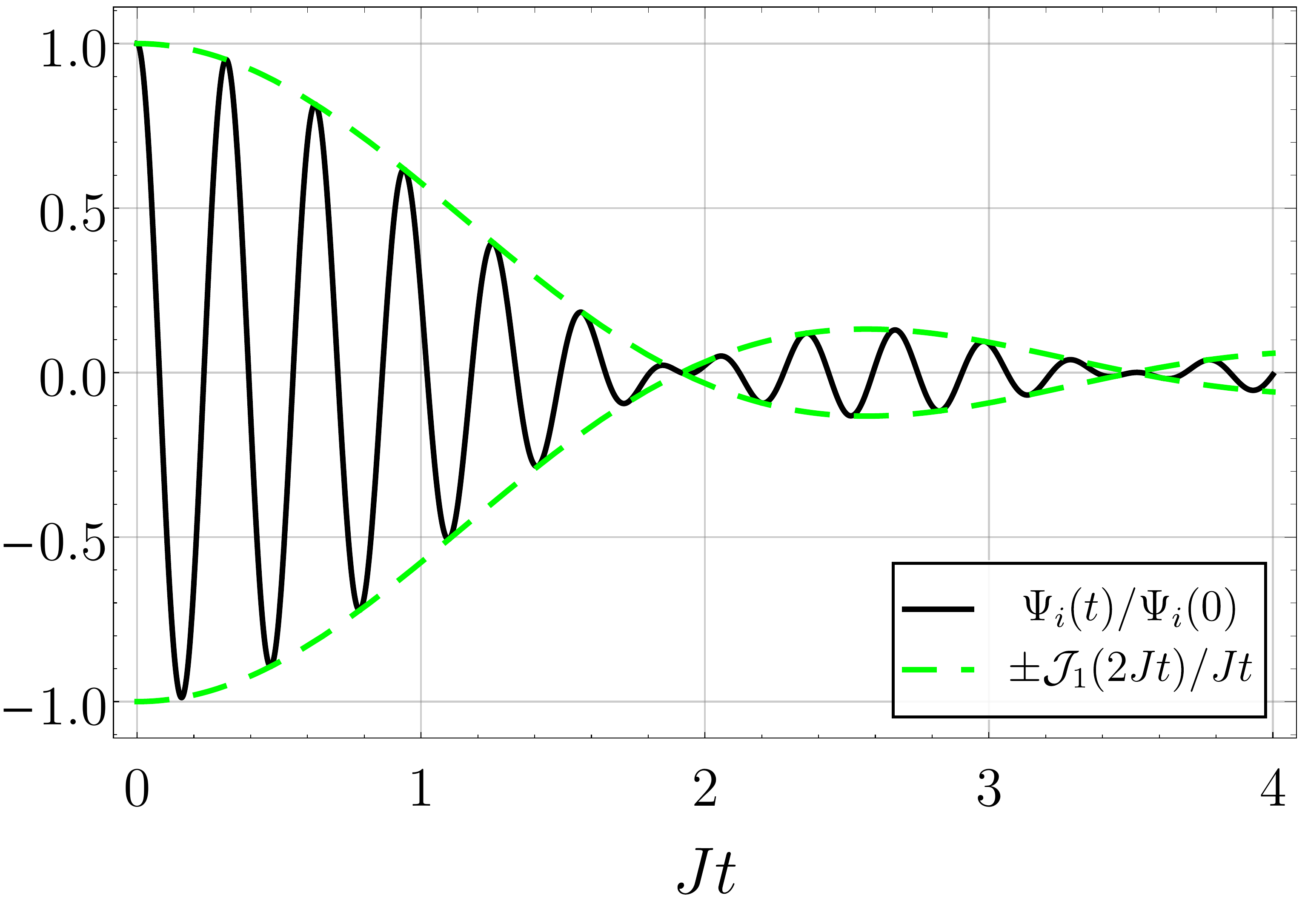}

\caption{\label{fig:LRT_RelaxationFunction}
Paramagnetic relaxation function of Eq.~\eqref{eq:ParamagneticRelaxFunction} for $\field_i = 10J$, together with its Bessel envelope.
}

\end{figure}

To test this optimization strategy with the relaxation function of an isolated system, we once again employ the TI chain.
Its relaxation function $\Psi_i(t)$ is given in Eq.~\eqref{eq:TIRelaxFunction} of appendix~\ref{sec:TIchain} when the chain is initially prepared in its ground state.
Deep in the paramagnetic phase ($\field_i \gg J$) and in the thermodynamic limit $N \to \infty$, we can swap sum by integral and approximate the quantities of Eq.~\eqref{eq:TIRelaxFunction} as
\[
\frac{\sin^2 k}{\epsilon_k^{3}(\field_i)} \approx \frac{\sin^2 k}{\field_i^3} \qquad \text{and} \qquad \epsilon_k(\field_i) \approx \field_i - J \cos k.
\]
This leads to
\begin{align}
\notag
\Psi_i(t) & \approx \frac{N J^2}{2\pi \field_i^3} \int_0^\pi \cos\left[ 2 (\field_i - J \cos k) t \right] \sin^2 k\, dk \\
\label{eq:ParamagneticRelaxFunction}
& = \frac{N J^2}{4\field_i^3} \frac{\mathcal J_1(2Jt)}{Jt} \cos(2\field_i t),
\end{align}
where $\mathcal J_n$ is the Bessel function of the first kind (the caligraphic $\mathcal J$ is used to avoid confusion with the coupling constant $J$).

\begin{figure*}

\subfloat[\label{fig:LRT_para_ptc_Jtau=2}]{\includegraphics[width=.33\textwidth]{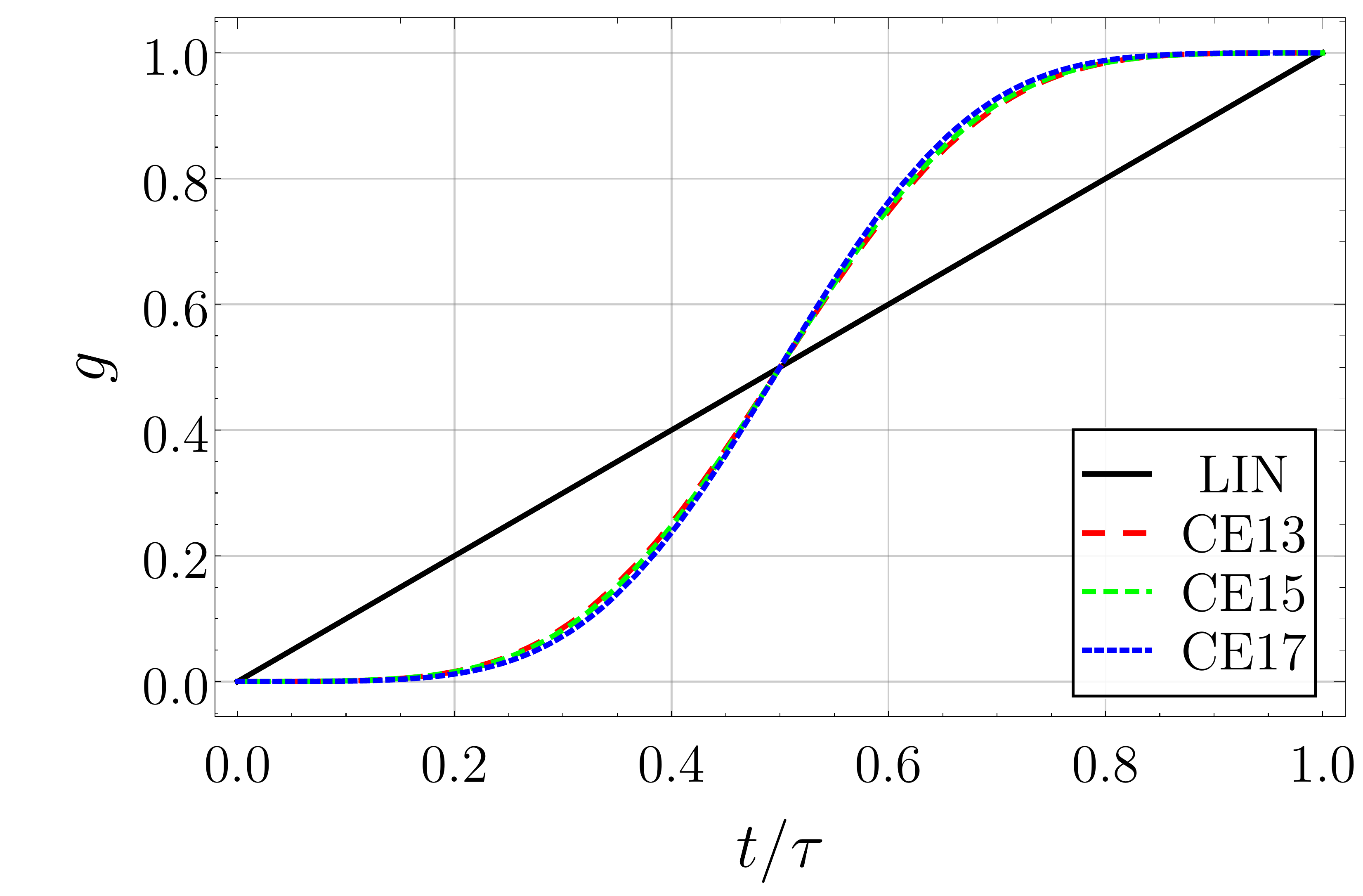}}
\subfloat[\label{fig:LRT_para_work_Jtau=2_Delta2}]{\includegraphics[width=.33\textwidth]{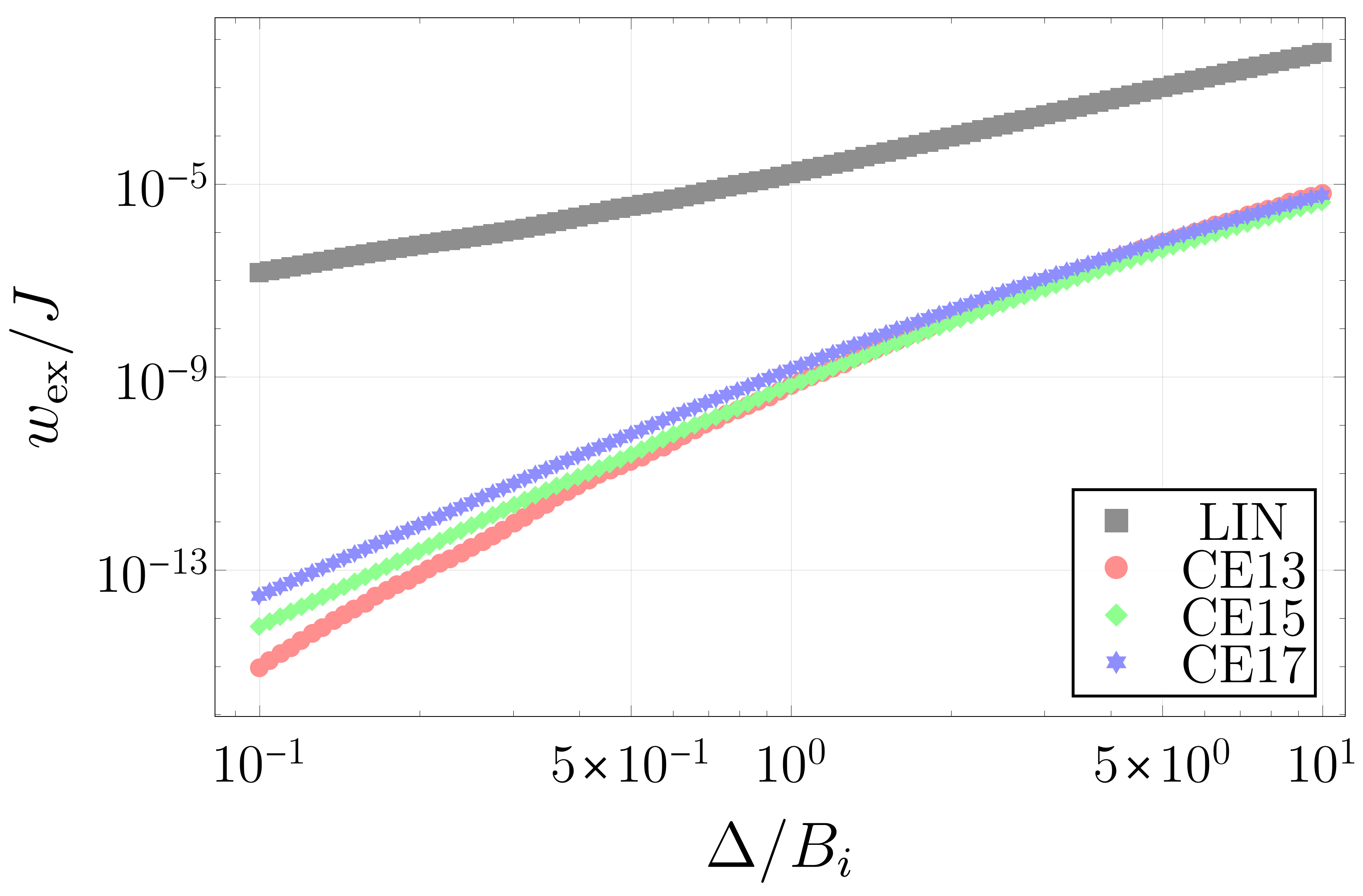}}
\subfloat[\label{fig:LRT_para_work_Jtau=2_Delta1}]{\includegraphics[width=.33\textwidth]{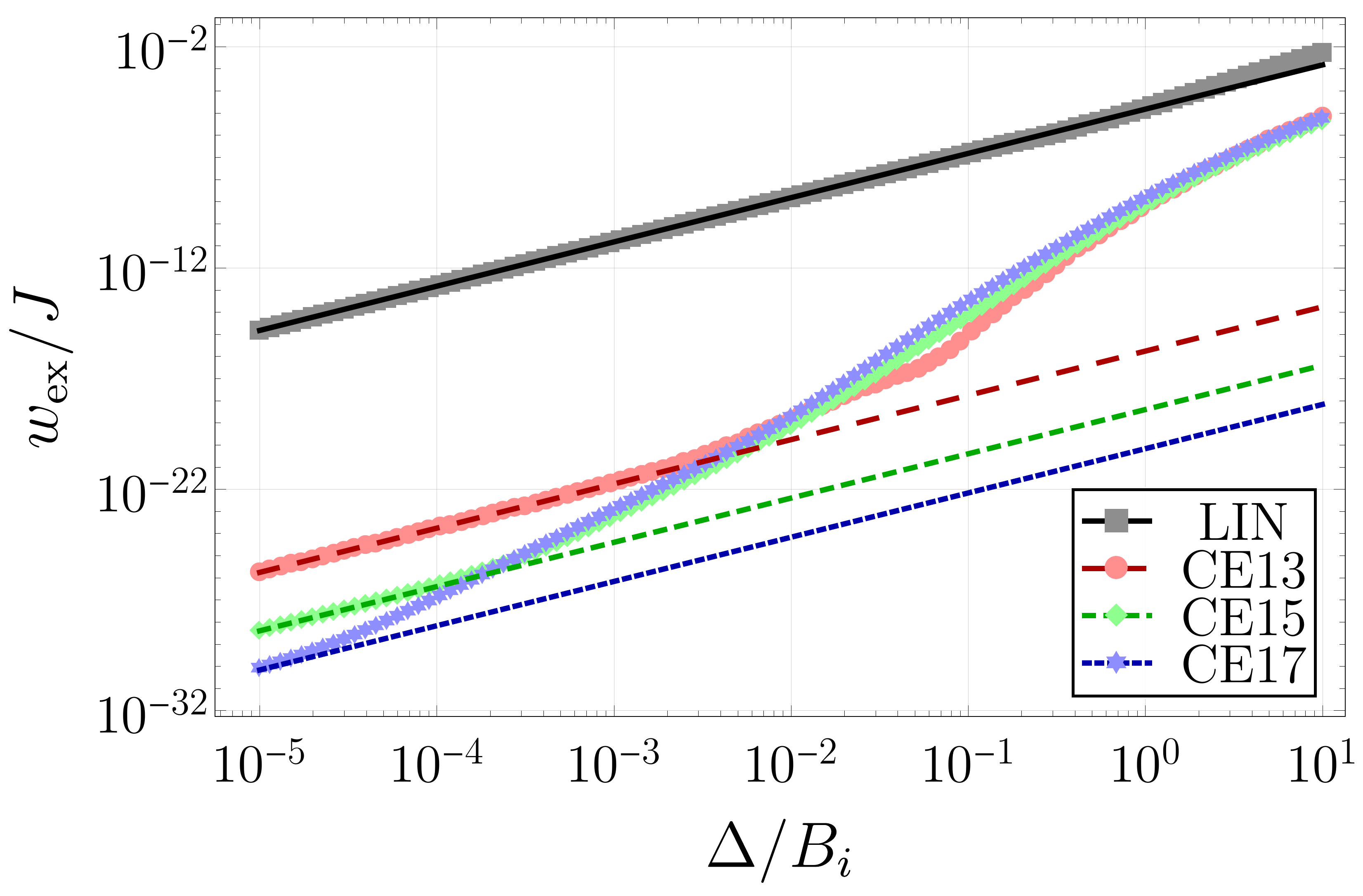}}

\subfloat[\label{fig:LRT_para_ptc_Jtau=1}]{\includegraphics[width=.33\textwidth]{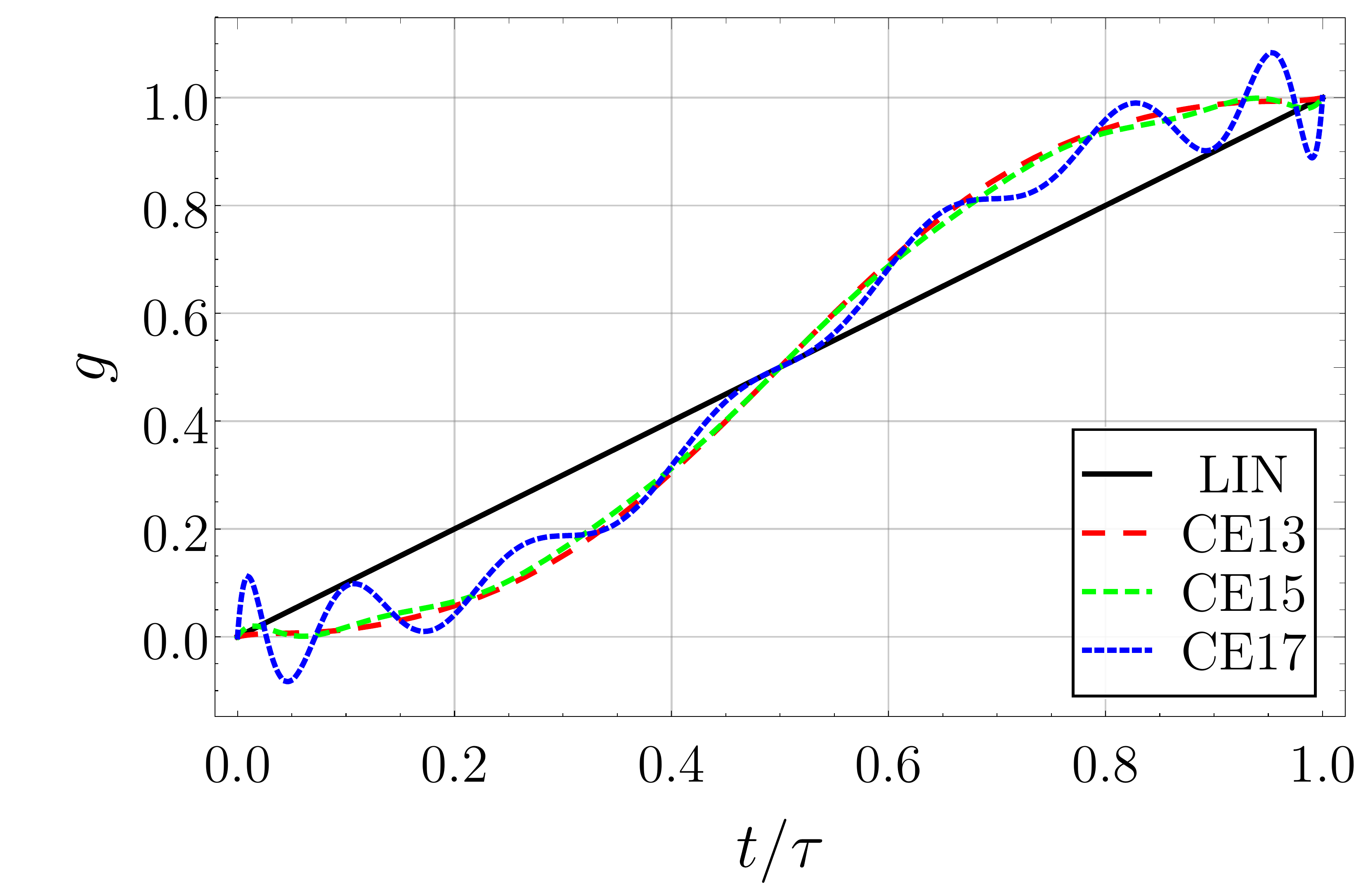}}
\subfloat[\label{fig:LRT_para_work_Jtau=1_Delta2}]{\includegraphics[width=.33\textwidth]{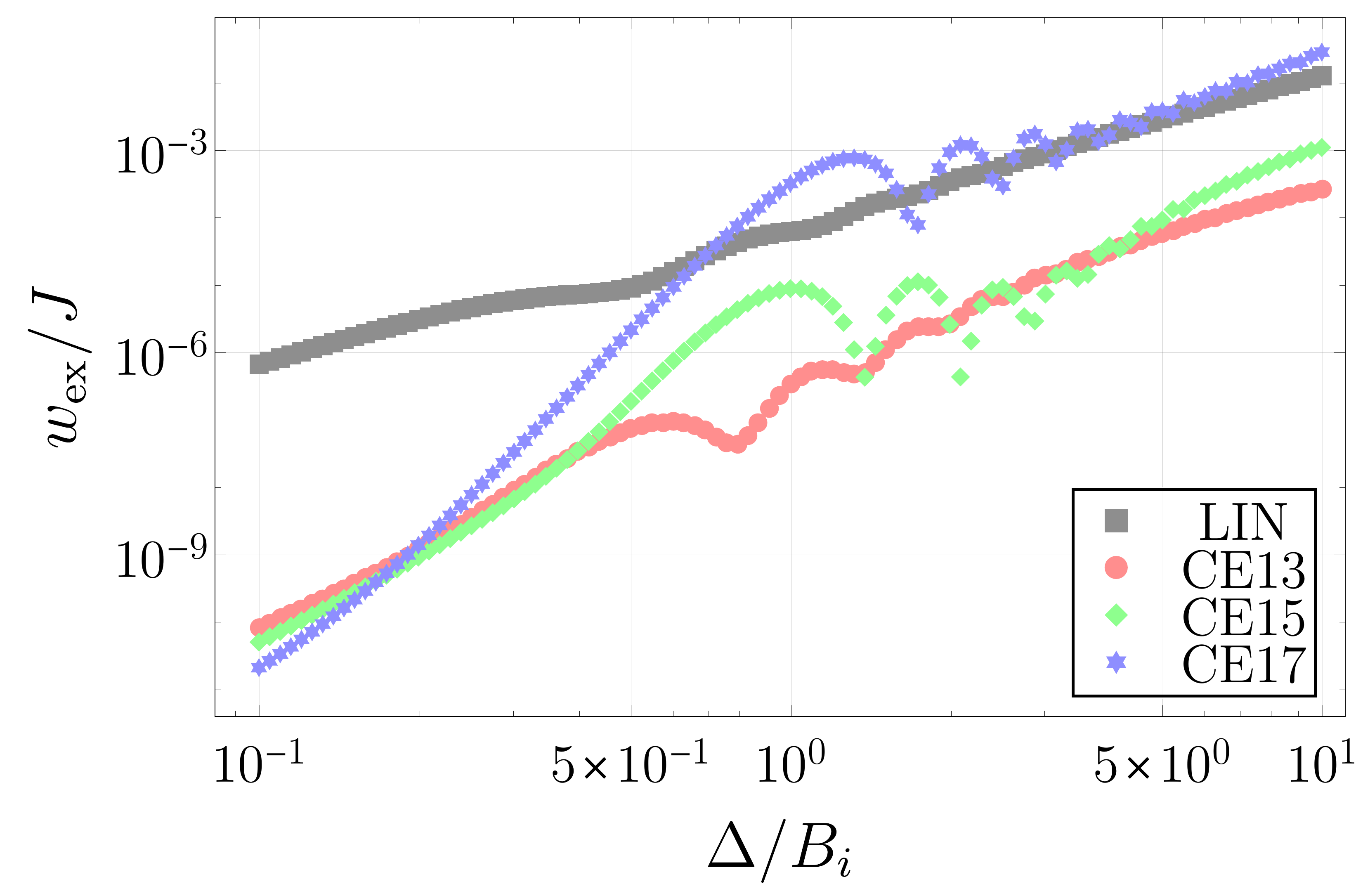}}
\subfloat[\label{fig:LRT_para_work_Jtau=1_Delta1}]{\includegraphics[width=.33\textwidth]{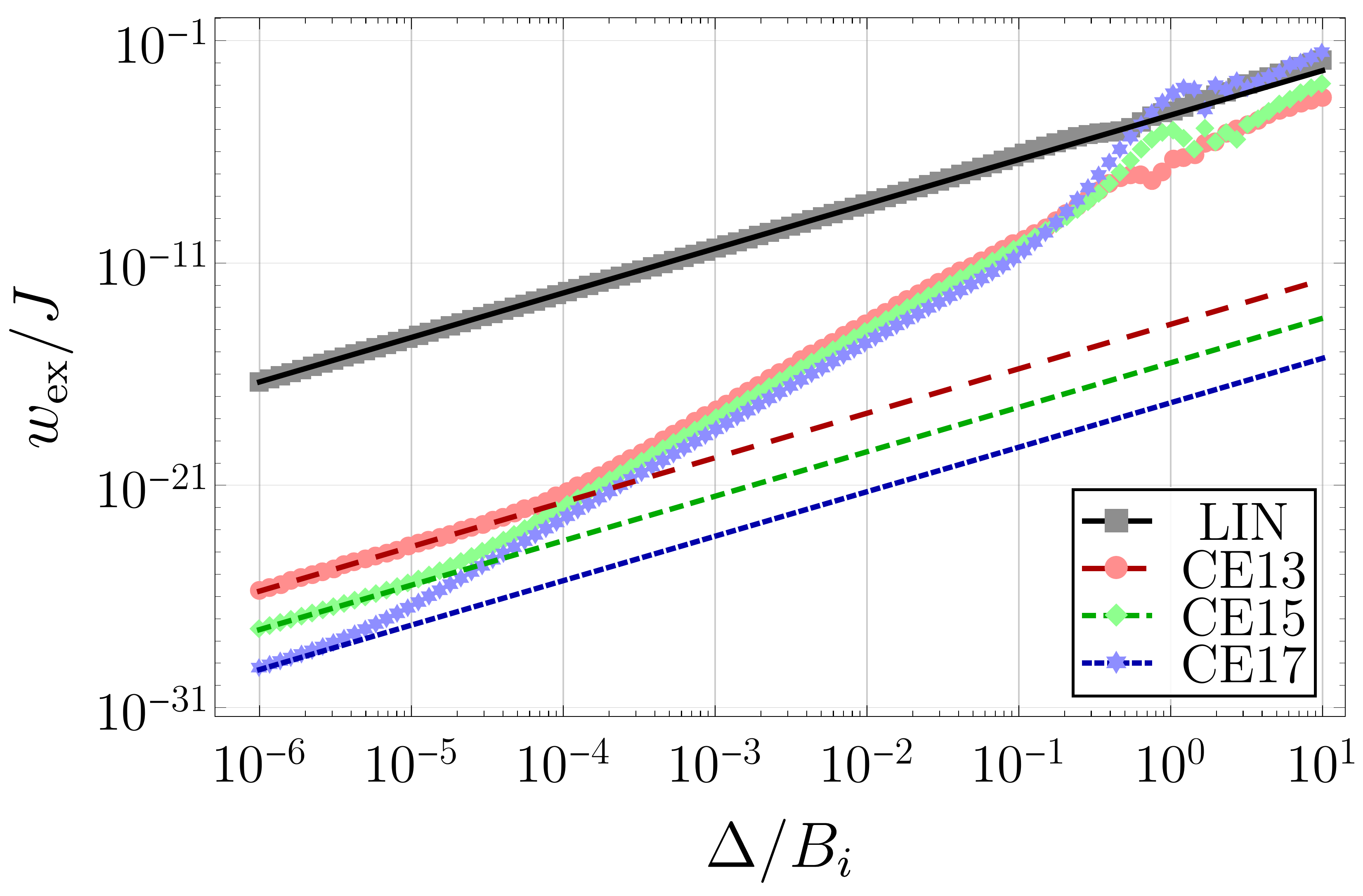}}

\subfloat[\label{fig:LRT_para_ptc_Jtau=0.5}]{\includegraphics[width=.33\textwidth]{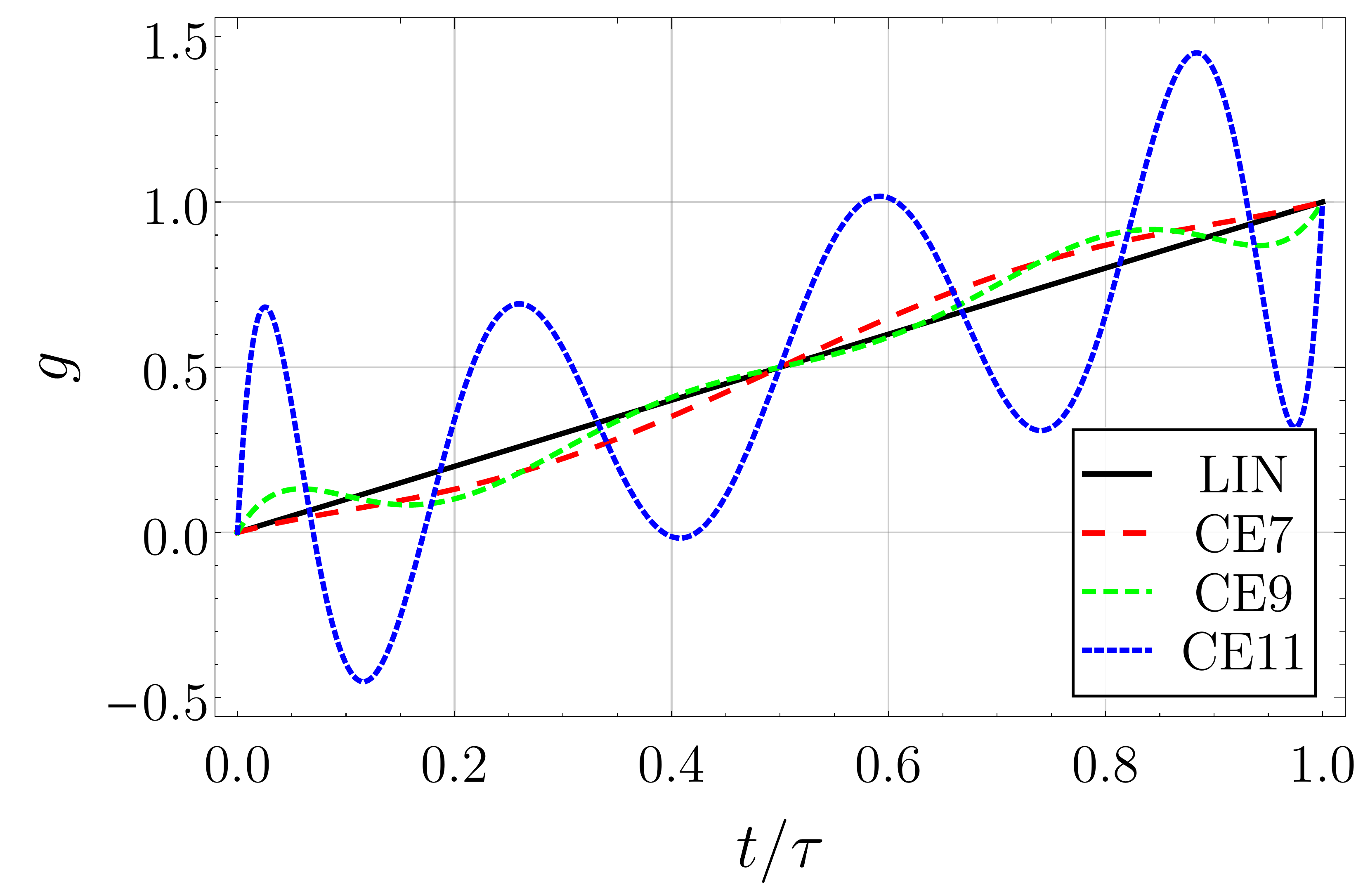}}
\subfloat[\label{fig:LRT_para_work_Jtau=0.5_Delta2}]{\includegraphics[width=.33\textwidth]{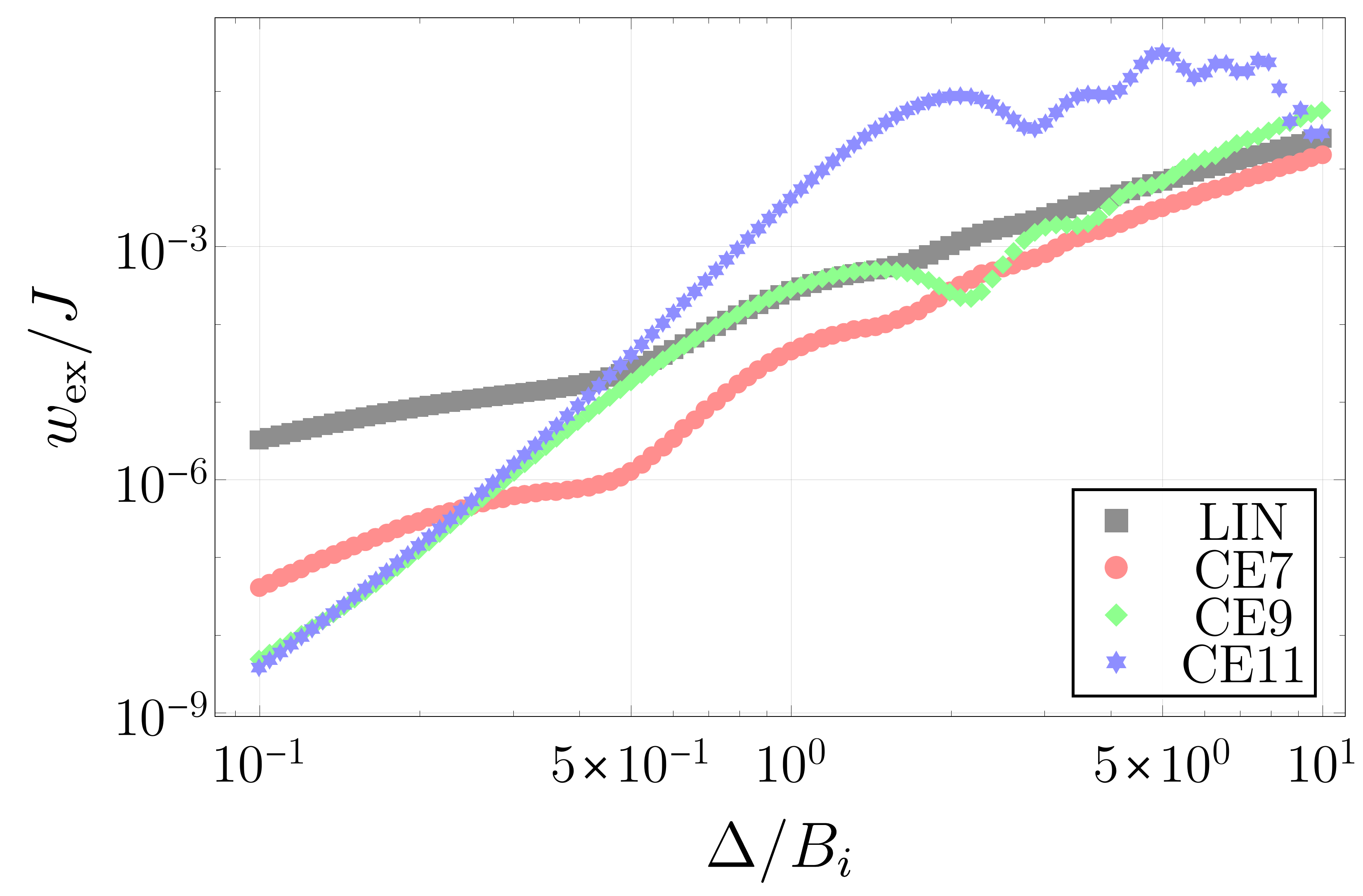}}
\subfloat[\label{fig:LRT_para_work_Jtau=0.5_Delta1}]{\includegraphics[width=.33\textwidth]{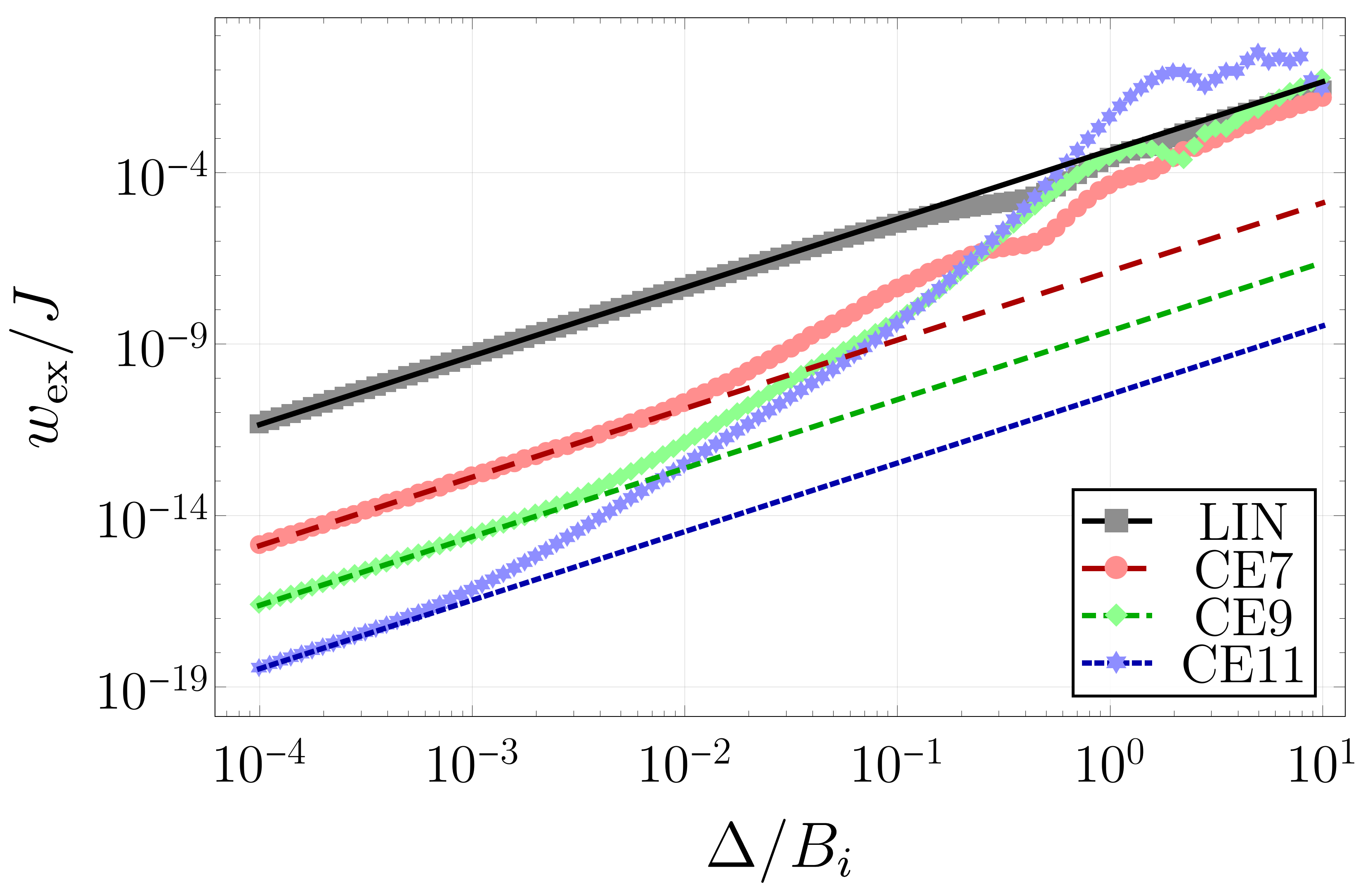}}

\caption{\label{fig:LRT_protocols}
Driving protocols and associated excess work per spin $w_\mathrm{ex} = W_\mathrm{ex}/N$ of Eq.~\eqref{eq:TIwork} in the TI chain with $N=100$ spins, for a paramagnetic process with $\field_i = 10J$.
The symbols in (b), (c), (e), (f), (h) and (i) depict numerical results obtained from the integration of the exact time-dependent dynamics.
In (c, f and i), the lines represent the LRT prediction.
The top plots (a, b and c) correspond to $J\tau=2$, the middle plots (d, e and f) correspond to $J\tau = 1$ and bottom plots (g, h and i) correspond to $J\tau = 0.5$.
LIN corresponds to the linear protocol, while CE$n$ corresponds to the Chebyshev expansion up to order $n$.
}

\end{figure*}

The function of Eq.~\eqref{eq:ParamagneticRelaxFunction} has a clear decay factor in $\mathcal J_1(2Jt) / Jt$, indicating a relaxation time scale of $J^{-1}$; and a pure oscillating factor in $\cos(2\field_i t)$, indicating an oscillation time scale of $\field_i^{-1}$ (see Fig.~\ref{fig:LRT_RelaxationFunction}).
From Eq.~\eqref{eq:TIDispersion}, it can be seen that the relevant energy gap of the system in the paramagnetic phase is $\field_i$.
Thus, in the TI chain, it is the oscillation time, not the relaxation time, that determines its closeness to adiabaticity.
The results using the relaxation function of Eq.~\eqref{eq:ParamagneticRelaxFunction} can be seen in Fig.~\ref{fig:LRT_protocols}.
The symbols in the plots for the excess work per spin depict the numerical results obtained from the integration of the exact time-dependent dynamics, while the lines represent the LRT prediction.

Figure~\ref{fig:LRT_para_ptc_Jtau=2} shows the linear protocol and three protocols obtained from the Chebyshev expansion for $J\tau = 2$.
Somewhat surprisingly, these protocols look identical to the BCM protocols we presented in the context of slow processes.
Indeed, a glance at Fig.\ref{fig:APT_para_work_tau} reveals that for this process duration, all of the protocols considered in that plot already agree with the APT prediction.
This indicates that $J\tau = 2$ can be considered slow, and the minimization of the LRT functional exploits this fact to generate protocols that follow the BCM guideline.
However, note that for realistic values of $\Delta$, i.e., $10^{-1} \leq \Delta/\field_i \leq 10^{1}$, increasing the number of modes worsens the performance, as can be seen on Fig.~\ref{fig:LRT_para_work_Jtau=2_Delta2}.
The work only decreases with increasing number of modes for much smaller values of $\Delta$ (see Fig.~\ref{fig:LRT_para_work_Jtau=2_Delta1}), closer to the point where LRT starts agreeing with the numerical data.
In any case, all three CE protocols outperform the linear one for the entire $\Delta$ range of Fig.~\ref{fig:LRT_para_work_Jtau=2_Delta1}.

The situation is considerably different for $J\tau = 1$.
Figure~\ref{fig:LRT_para_ptc_Jtau=1} compares the linear protocol once again with three Chebyshev expansions, with the same number of modes considered in the previous case.
In this case, we start seeing deviations from BCM protocols, but only when using 15 modes or more.
Essentially, the LRT functional cannot output better protocols than BCM ones when given a small number of degrees of freedom to optimize.
When it has enough degrees to work with, it generates oscillating protocols that outperform BCM-like protocols while escaping APT description.
Performance-wise, Fig.~\ref{fig:LRT_para_work_Jtau=1_Delta2} shows that increasing the number of modes does seem to decrease the excess work around $\Delta/\field_i = 10^{-1}$, but not beyond this point.
In fact, the oscillating protocol CE17 is outclassed even by the linear protocol for $\Delta/\field_i$ values as low as $10^{0}$, demonstrating that such oscillating protocols are only reliable in the LRT regime $\Delta/\field_i \ll 1$.

For $J\tau = 1$, increasing the number of modes past $17$ starts giving non-realistic oscillatory protocols, with amplitudes several orders of magnitude higher than the endpoint.
The maximum number of modes that still gives realistic protocols in fact decreases with decreasing $\tau$.
Figure~\ref{fig:LRT_para_ptc_Jtau=0.5} shows LIN and three CE protocols, the highest number of modes used being $11$, for $J\tau = 0.5$.
CE11 already shows considerable oscillations --- the protocols for higher number of modes are impractical.
Once again, the performance of these oscillatory protocols is only good for small values of $\Delta$: Fig.~\ref{fig:LRT_para_work_Jtau=0.5_Delta2} reveals that, when compared to CE9, CE11 has a marginal advantage at best and a large disadvantage at worst.

The trend of the $J\tau = 0.5$ (Fig.~\ref{fig:LRT_para_ptc_Jtau=0.5}) and $J\tau = 1$ (Fig.~\ref{fig:LRT_para_ptc_Jtau=1}) cases studied seems to indicate that, using a high enough number of modes, oscillating protocols can be generated from the LRT functional for any value of $\tau$.
However, we have not found oscillating protocols for $J\tau = 2$ (Fig.~\ref{fig:LRT_para_ptc_Jtau=2}) for a number of modes of up to 23, where numerical errors started becoming a factor.

\begin{figure*}

\subfloat[\label{fig:LRT_para_work_Delta=1J}]{\includegraphics[width=.33\textwidth]{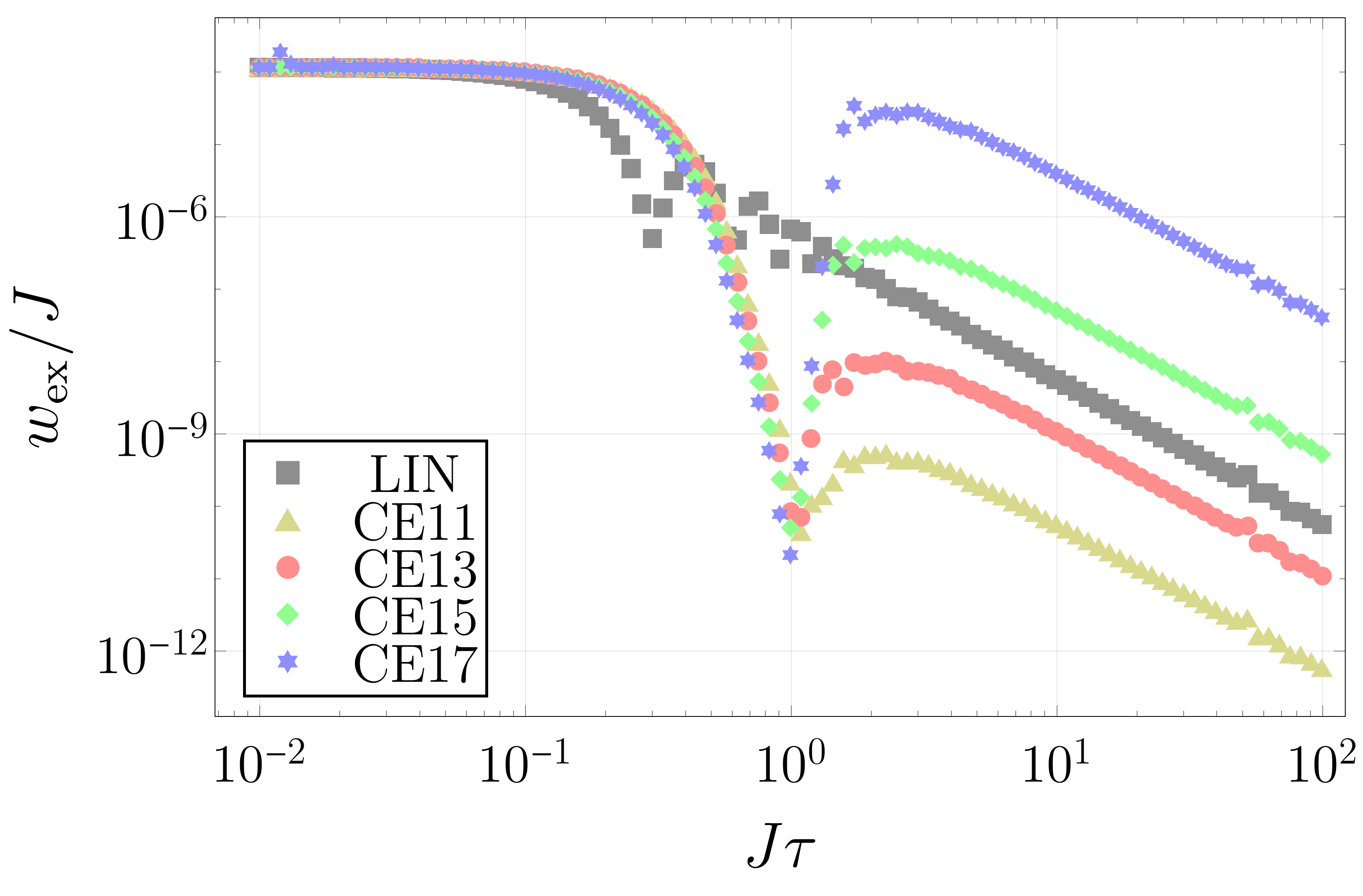}}
\subfloat[\label{fig:LRT_para_work_Delta=10J}]{\includegraphics[width=.33\textwidth]{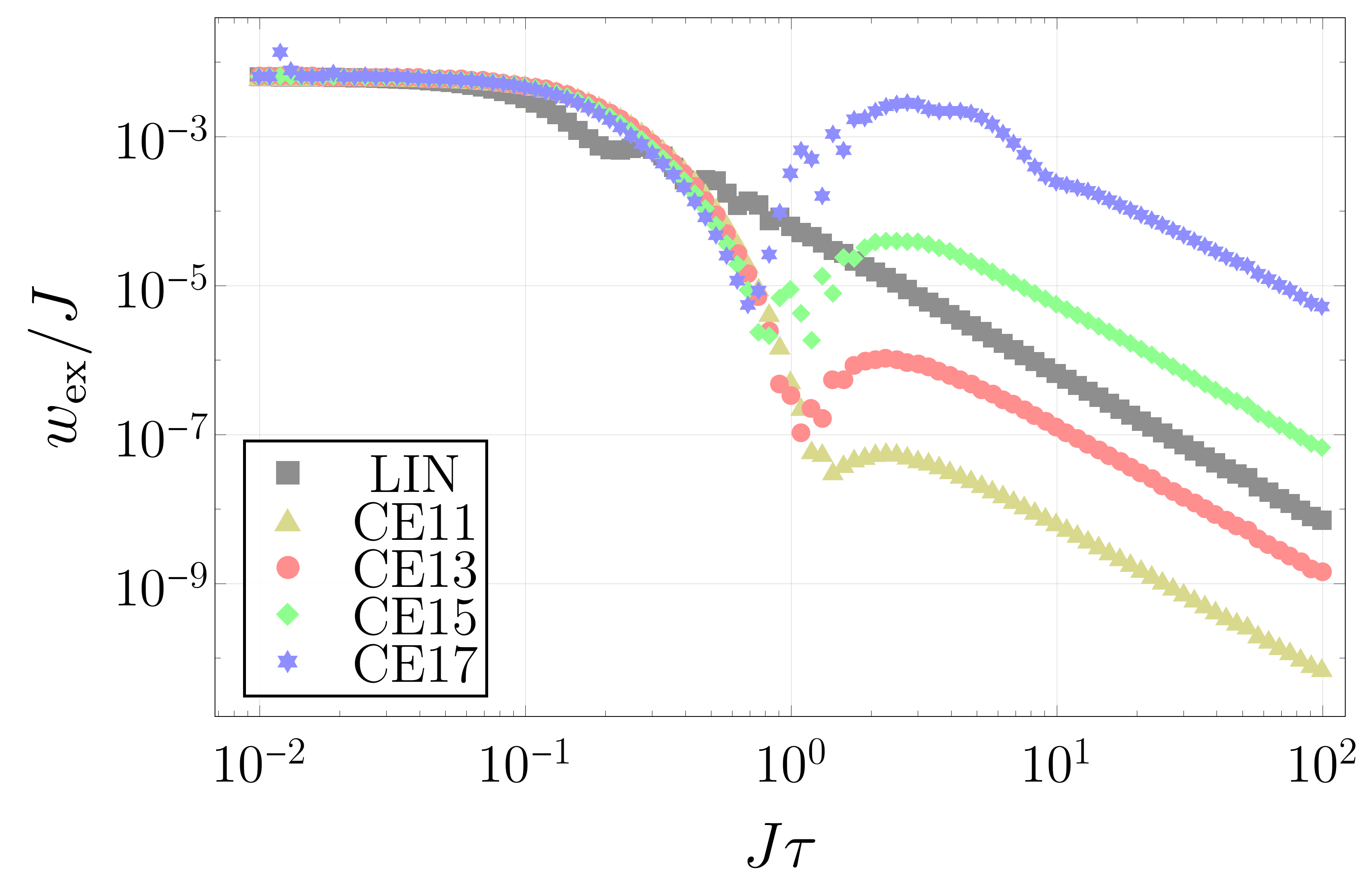}}
\subfloat[\label{fig:LRT_para_work_Delta=100J}]{\includegraphics[width=.33\textwidth]{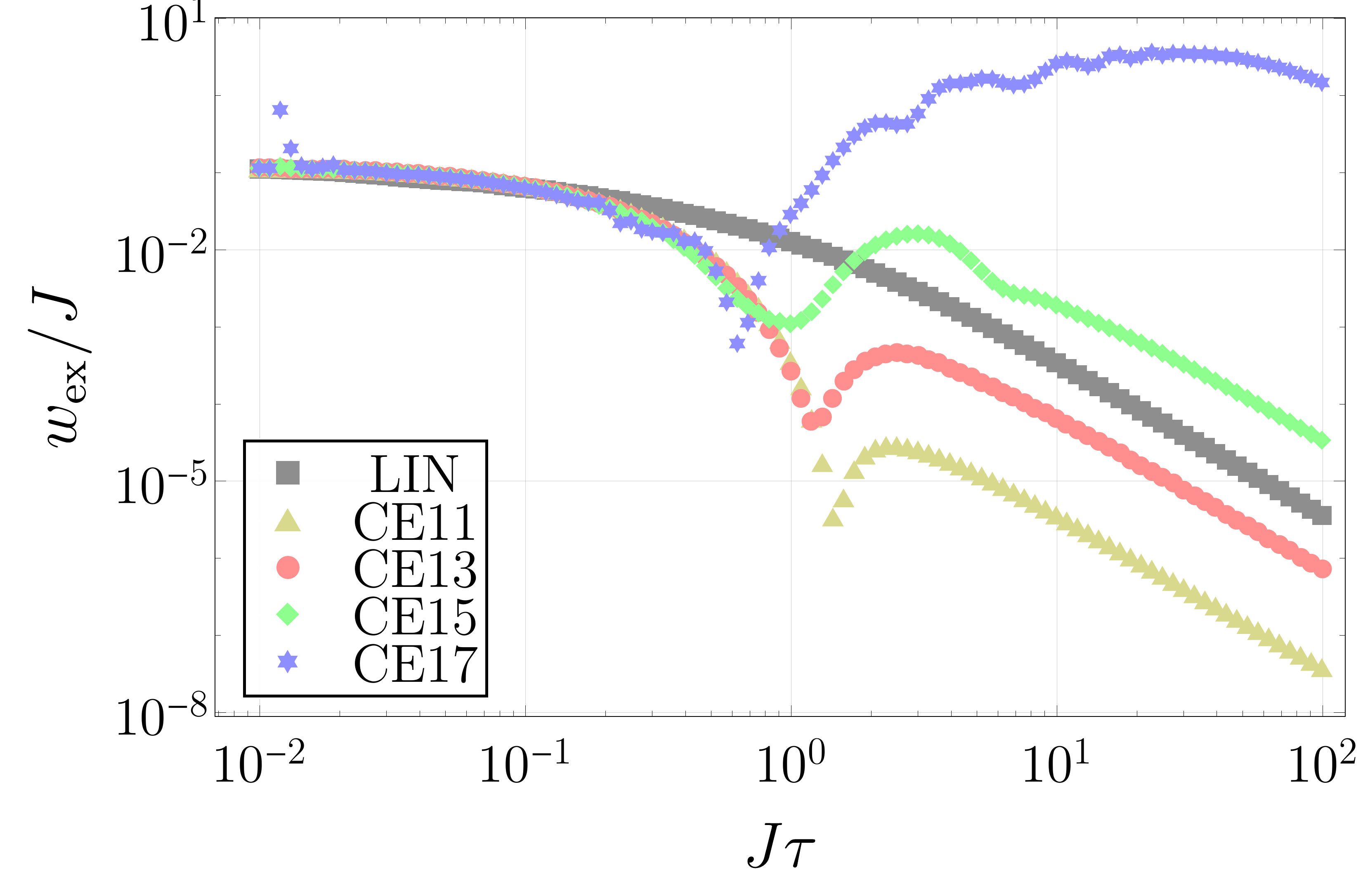}}

\caption{\label{fig:LRT_work_tau}
Excess work per spin $w_\mathrm{ex} = W_\mathrm{ex}/N$ of Eq.~\eqref{eq:TIwork} for the protocols of Fig.~\ref{fig:LRT_para_ptc_Jtau=1} and \ref{fig:LRTBCM_para_ptc_Delta=1J} in a paramagnetic process in the TI chain with $N=100$ spins and $\field_i = 10J$, obtained from the integration of the exact time-dependent dynamics.
LIN corresponds to the linear protocol, while CE$n$ corresponds to the Chebyshev expansion up to order $n$.
The total variation of the field used was (a) $\Delta = J$, (b) $\Delta = 10J$ and (c) $\Delta = 100J$.
}

\end{figure*}

\begin{figure*}

\subfloat[\label{fig:LRTBCM_para_ptc_Delta=1J}]{\includegraphics[width=.33\textwidth]{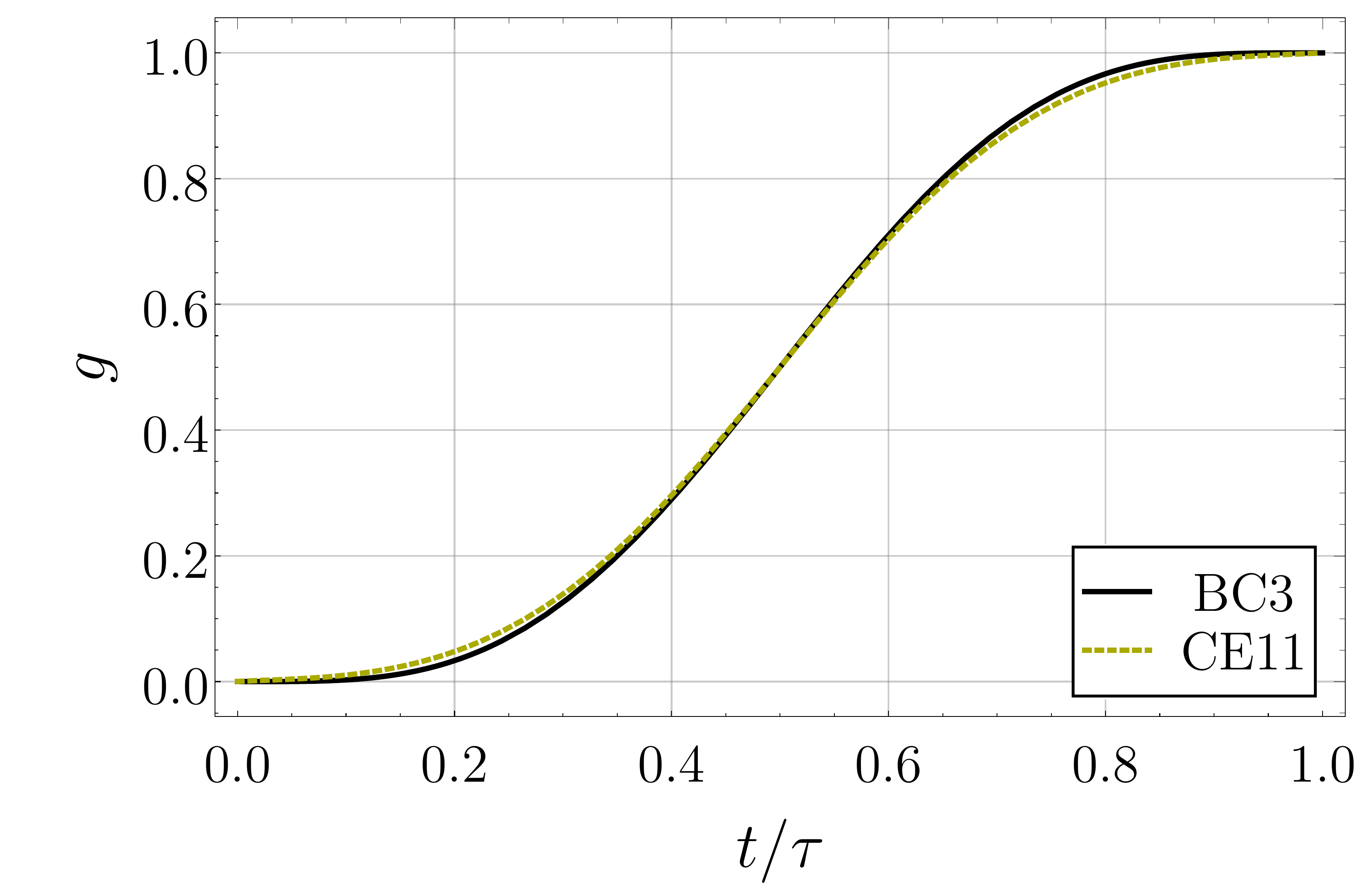}}
\subfloat[\label{fig:LRTBCM_para_Dptc_Delta=1J}]{\includegraphics[width=.33\textwidth]{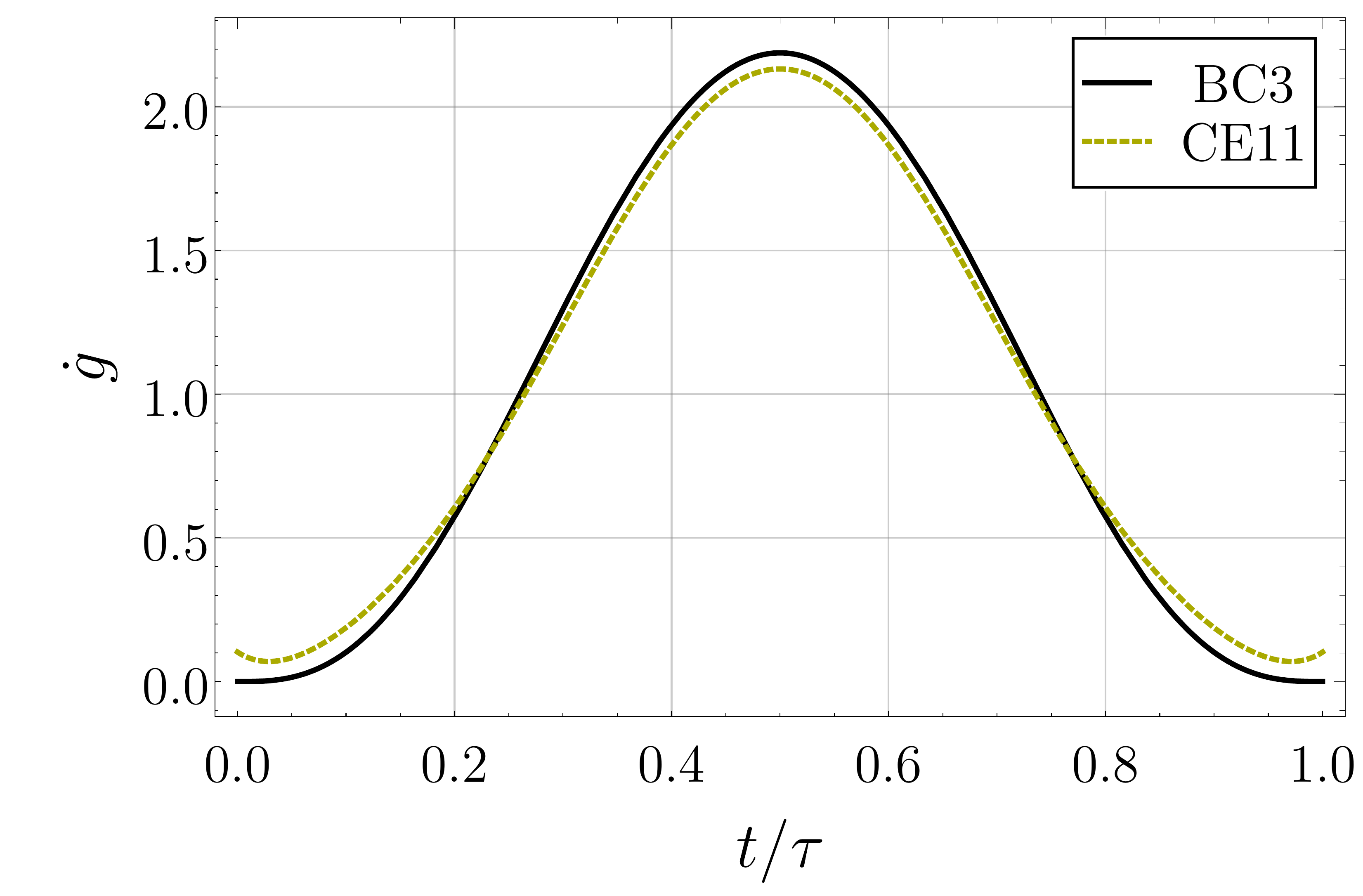}}
\subfloat[\label{fig:LRTBCM_para_work_Delta=1J}]{\includegraphics[width=.33\textwidth]{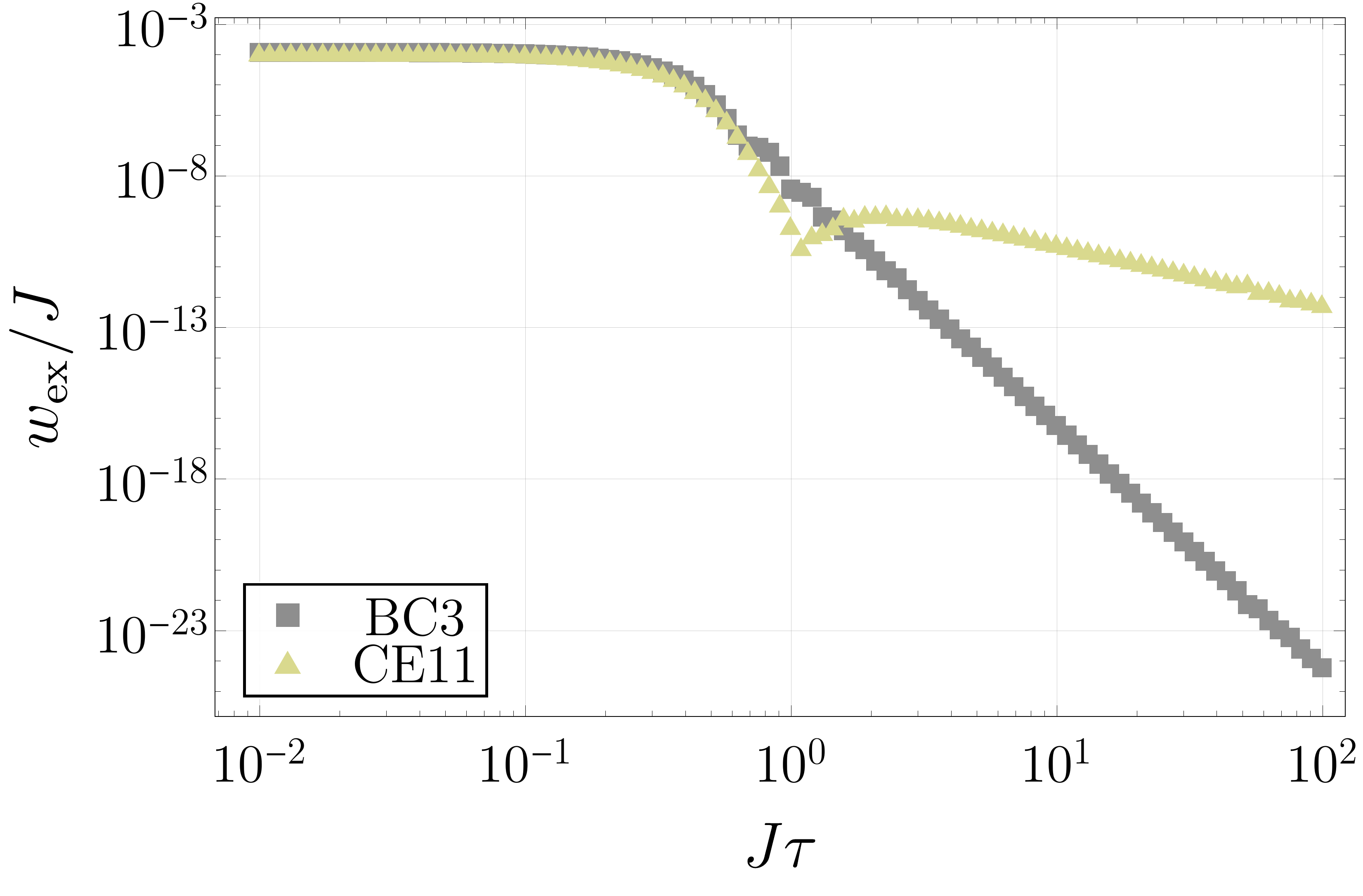}}

\caption{\label{fig:LRTBCM}
Driving protocols and associated excess work per spin $w_\mathrm{ex} = W_\mathrm{ex}/N$ of Eq.~\eqref{eq:TIwork} in the TI chain with $N=100$ spins, for a paramagnetic process with $\field_i = 10J$.
(a) Protocols BC3 (from Eq.~\eqref{eq:BCMprotocols} with $r=3$) and CE11 (from the LRT functional with 11 Chebyshev modes).
(b) First time derivatives of the same protocols.
(c) Excess work vs $\tau$ for the same protocols.
It was obtained from the numerical integration of the exact time-dependent dynamics.
}

\end{figure*}

Figure~\ref{fig:LRT_work_tau} shows the excess work per spin (obtained numerically) vs $\tau$ for the protocols shown in Fig.~\ref{fig:LRT_para_ptc_Jtau=1} and CE11 shown in Fig.~\ref{fig:LRTBCM_para_ptc_Delta=1J}.
These plots demonstrate how specific are the protocols obtained from the LRT functional for the values of $\tau$ used to generate them.
Figure~\ref{fig:LRT_para_work_Delta=1J} was made with $\Delta/\field_i = 0.1$, and it shows that all four CE protocols have sharp valleys centered at $J\tau = 1$, which is exactly the value of the process duration they were designed to optimize.
For values of $\tau$ outside this valley, the CE15 and CE17 protocols are outperformed by LIN.
CE11 and CE13, on the other hand, maintain better performance than LIN for higher values of $\tau$, a consequence of their BCM-like appearance.
As $\Delta$ is increased, the valleys become dispersed and less profound, as can be seen in Fig.~\ref{fig:LRT_para_work_Delta=10J} for $\Delta/\field_i = 1$.
When the system is far from LRT description, as for $\Delta/\field_i = 10$ in Fig.~\ref{fig:LRT_para_work_Delta=100J}, the valleys still exist, but the protocols CE15 and CE17 lose any advantage they had --- another sign that the oscillatory CE protocols only guarantee a decrease in the excess work when one stays in the weak regime, as mentioned before.

Note that, in accordance to what we discussed in the context of slow processes, for every protocol considered in Fig.~\ref{fig:LRT_work_tau}, the excess work decays as $\tau^{-2}$ for large $\tau$.
This includes CE11 and CE13, which one would naively predict to have steeper decays, as expected from BCM protocols.
The reason for this is the fact that CE11, for instance, is not exactly a BCM protocol --- after all, unlike Eq.~\eqref{eq:BCMprotocols}, it has to accommodate for the specific properties of the system, imported from the relaxation function.
Figure~\ref{fig:LRTBCM} compares CE11 with BC3, obtained from Eq.~\eqref{eq:BCMprotocols} with $r = 3$.
They are almost indistinguishable from the point of view of Fig.~\ref{fig:LRTBCM_para_ptc_Delta=1J}, but Fig.~\ref{fig:LRTBCM_para_Dptc_Delta=1J} reveals that CE11 has finite first derivatives at the endpoints, which means APT gives $W_\mathrm{ex} \sim \tau^{-2}$.
Conversely, BC3 has a considerably steeper decay of $\tau^{-8}$ in the APT regime, as can be seen in Fig.~\ref{fig:LRTBCM_para_work_Delta=1J}.
Nevertheless, such small derivatives are the reason why CE11 consistently beats LIN in Fig.~\ref{fig:LRT_work_tau}.

The absence of the universal $\tau^{-1}$ decay for large $\tau$ is hence corroborated using another perturbation theory.
We emphasize that the analytical results of this section considered the thermodynamic limit of the Ising chain through expressions such as \eqref{eq:ParamagneticRelaxFunction}.
Hence, the $\tau^{-1}$ scaling does not seem to be related to this asymptotic limit.
The agreement between both perturbation theories used in this paper in the appropriate \emph{weak and slow} regime is elucidated in appendix~\ref{sec:APTandLRTagree}.

In passing, we remark that the expression for the relaxation function of Eq.~\eqref{eq:TIRelaxFunction}, deep in the ferromagnetic phase ($\field \ll J$) and in the thermodynamic limit, is approximated by
\begin{equation} \label{eq:FerromagneticRelaxFunction}
\Psi_i(t) = \frac{N}{4J} \frac{\mathcal J_1(2\field_i t)}{\field_i t} \cos(2J t).
\end{equation}
Disregarding the constant prefactor, Eq.~\eqref{eq:FerromagneticRelaxFunction} is very similar to Eq.~\eqref{eq:ParamagneticRelaxFunction}, only differing by the substitution $\field_i \leftrightarrow J$.
This is related to the Kramers--Wannier duality between the two phases of the system \cite{Radicevic2018}.
Consequently, the LRT results for the ferromagnetic phase are the same as those for the paramagnetic phase, given suitable values of $\field_i$ and $\tau$.
For instance, the protocols obtained from the LRT functional for $\field_i = 0.1J$ and $J\tau = 10$ are identical to the protocols obtained for $\field_i = 10J$ and $J\tau = 1$, given in Fig.~\ref{fig:LRT_para_ptc_Jtau=1}.

\section{Nonintegrable systems \label{sec:NonIntegrable}}

The optimization strategies presented and discussed in this paper are naturally written in terms of the eigen-quantities of the Hamiltonian.
Yet, unlike the transverse-field Ising Hamiltonian of Eq.~\eqref{eq:TIHamiltonian}, the vast majority of Hamiltonians cannot be diagonalized exactly, i.e., most systems are nonintegrable.
When the system is numerically integrable, the strategies contained herein are readily available.
For the application of the Chebyshev expansion method, the eigen-equation need only be numerically solved at $\lambda_i$.
However, the eigen-equation must be solved for every $\lambda$ value traversed in the defined process for the application of the FQA method, since it requires knowledge of how the spectrum changes with $\lambda$.

The optimization strategies contained herein can also be applied in experimental situations where the linear response of the system can be determined through measurements.
To elaborate, the methods can be used if the generalized force $F$ --- which is $\lambda$-independent if the external parameter is linearly coupled to the system --- is known and if the response function can be determined from the correlation function
\begin{equation} \label{eq:AutoCorrelation}
\Phi(\lambda;t-t') = \frac{1}{i\hbar} \tr{ \rho^{(0)}(\lambda) [ F(\lambda;t) , F(\lambda;t') ] },
\end{equation}
where
\begin{equation} \label{eq:DensityMatrixAdiabatic}
\rho^{(0)}(\lambda) = \sum_n p_n \ket{n(\lambda)} \bra{n(\lambda)}
\end{equation}
is the (equilibrium) density matrix in the adiabatic limit, $[\cdot,\cdot]$ denotes the commutator and
\begin{equation} \label{eq:ForceInteractionPicture}
F(\lambda;t) = e^{-(t-t_i)H(\lambda)/i\hbar} F(\lambda) e^{(t-t_i)H(\lambda)/i\hbar}
\end{equation}
is the generalized force in the instantaneous interaction picture of $H(\lambda)$.

Note that the trace in Eq.~\eqref{eq:AutoCorrelation} is an equilibrium average, since the state $\rho^{(0)}$ is diagonal.
The success of linear response theory is largely based on the measurability of this average.
For instance, with an initially canonical distribution, one needs to determine $\Phi_i(t) = \Phi(\lambda_i;t)$ and calculate $\Psi_i(t)$ to apply the Chebyshev expansion method.
On the other hand, to determine fast quasi-adiabatic protocols, one first rewrites Eq.~\eqref{eq:FQAAltDifferentialEquation} as
\begin{equation} \label{eq:FQAAltDifferentialEquation2}
\frac{\left\vert \Upsilon(\lambda;0) \right\vert}{2} \dot\lambda^2(t) = c \left\vert \tr{ \rho^{(0)}(\lambda) H(\lambda) } \right\vert
\end{equation}
where $\Upsilon(\lambda;t)$ is a function described in appendix~\ref{sec:APTandLRTagree}, while the right-hand side contains the average energy in the adiabatic limit.
Then one needs to determine the average in Eq.~\eqref{eq:AutoCorrelation} for every $\lambda$ in the process, which might not be an easy task since the state of the system may not maintain canonical form throughout the evolution, as discussed in Sec.~\ref{sec:ExcessWork}.
If this can be done, one then calculates $\Upsilon$ from $\Phi(\lambda;t) = - \pd_t^3 \Upsilon(\lambda;t)$ and solves the differential equation~\eqref{eq:FQAAltDifferentialEquation2}.

\section{Open systems \label{sec:OpenSystems}}

In this section, we briefly consider the implications of our analysis to open quantum systems.
We stress that at no point during our derivation of the excess work in slow processes (Sec.~\ref{sec:SlowProcesses}) did we assume that the external parameter $\lambda$ interacts with the entire isolated system.
Thus, in a gapped bipartite system (call it $U$), if $\lambda$ acts only on one of the parts (call it $S$), the excess work exerted on $U$ still decays asymptotically as (at least) $\tau^{-2}$ for slowly-varying processes (we also assume the total energy of $U$ is well below its gap).
However, since the force applied on $S$ must also be the total force applied on $U$ (there are no other external influences), it follows that \emph{the work exerted on the open system $S$ is equal to the work exerted on the closed system $U$}.
Consequently, through a Hamiltonian description of a closed and gapped quantum system $U$ that includes the open quantum system $S$, we can see that the excess work done on open systems also scales as $\tau^{-2}$ in slow processes. 

The above result can be made precise with the operational definitions of average work used in this paper.
The Hamiltonian of the isolated system $U$, whose control parameter couples only with the part $S$, can always be written as
\begin{equation} \label{eq:TotalHamiltonian}
H_U(\lambda) = H_S(\lambda) + H_I + H_R,
\end{equation}
where $H_S(\lambda)$ is the Hamiltonian of the open system we manipulate (thus dependent on $\lambda$), $H_R$ is the Hamiltonian of the rest $R$ (thus independent of $\lambda$) and $H_I$ is the Hamiltonian of the interaction between $S$ and $R$ (also independent of $\lambda$).
Now, using Eq.~\eqref{eq:PowerWork} to calculate the work exerted on $U$, we see that
\begin{align}
\notag
W_U & = \int_{t_i}^{t_f} \dot\lambda(t) \text{Tr}_U\left\{ \rho_U(t) \frac{\pd H_U(\lambda)}{\pd\lambda} \right\} \\
\notag
& = \int_{t_i}^{t_f} \dot\lambda(t) \text{Tr}_S \text{Tr}_R \left\{ \rho_U(t) \frac{\pd H_S(\lambda)}{\pd\lambda} \right\} \\
\notag
& = \int_{t_i}^{t_f} \dot\lambda(t) \text{Tr}_S \left\{ \rho_S(t) \frac{\pd H_S(\lambda)}{\pd\lambda} \right\} \\
\label{eq:OpenSystemWork}
& = W_S,
\end{align}
where we used Eq.~\eqref{eq:TotalHamiltonian} to write $\pd_\lambda H_U(\lambda) = \pd_\lambda H_S(\lambda)$, the fact that the trace over the entire system can be decomposed into partial traces ($\text{Tr}_U = \text{Tr}_S \text{Tr}_R$) and $\rho_S(t) = \text{Tr}_R \rho_U(t)$ is the reduced density matrix of $S$.
In the last line of Eq.~\eqref{eq:OpenSystemWork}, we used Eq.~\eqref{eq:PowerWork} once again to identify the average work exerted on $S$.
Therefore, the general conclusions we obtained for the work in closed gapped quantum systems also applies to open quantum systems (assuming system plus environment is also gapped), including the absence of the asymptotic $\tau^{-1}$ behavior of the excess work in slowly varying processes.

We emphasize that the above arguments are rather general.
In particular, we make no assumption about the size of the system, nor do we assume specific features of the interaction between the open system and its environment.
However, we do assume that the initial state of the isolated system $U$ commutes with the initial Hamiltonian \eqref{eq:TotalHamiltonian}, a choice that exactly describes an initial thermodynamic equilibrium between $S$ and $R$.

\begin{figure}

\includegraphics[width=\columnwidth]{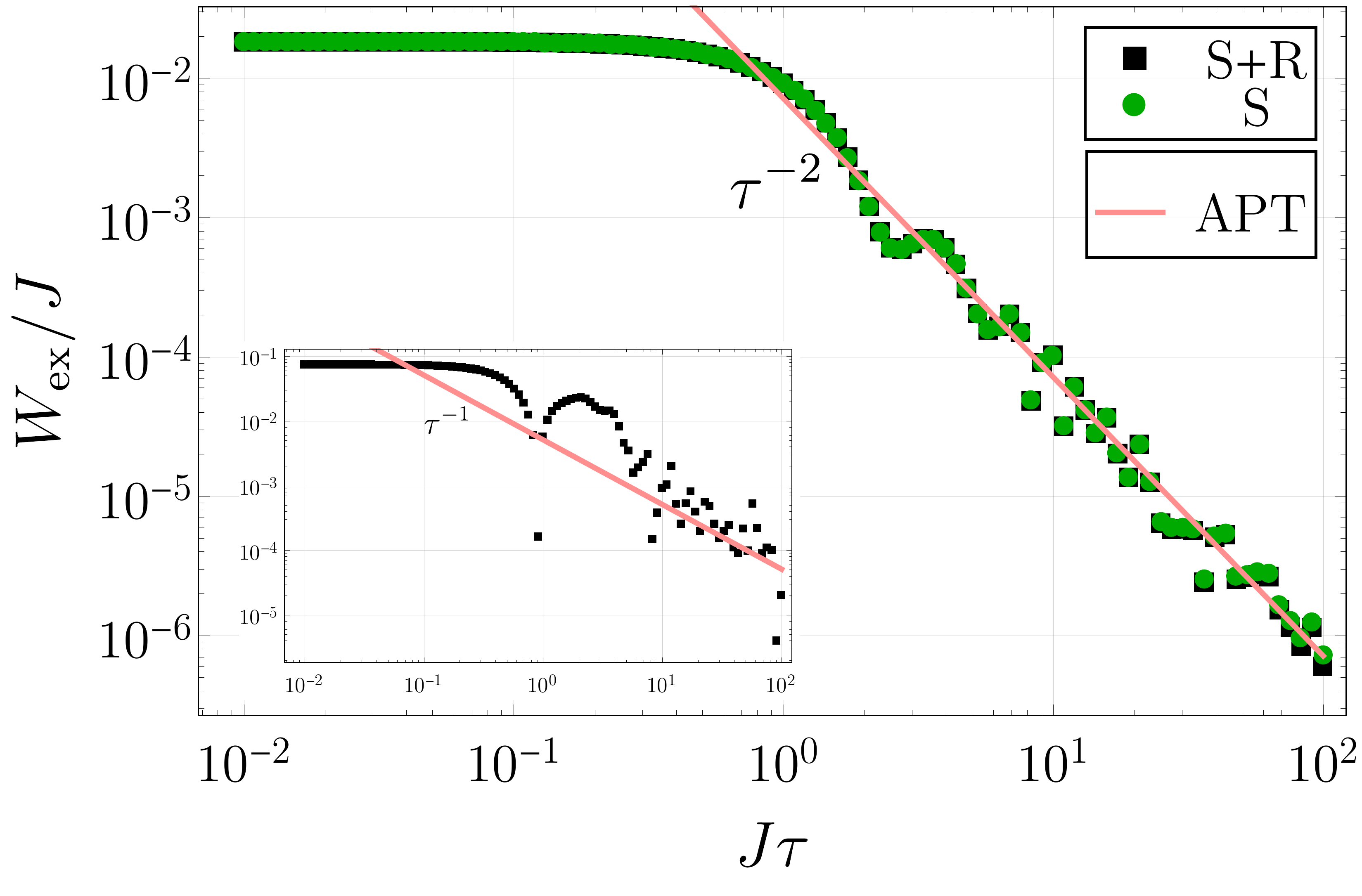}

\caption{\label{fig:3spins}
Excess work vs process duration for a 3-spin Ising chain calculated numerically from the exact dynamics.
The magnetic field applied on one of the spins was varied linearly in time from $2J$ to $3J$, while the magnetic fields in the other two spins were kept constant at $3J$.
S+R represents the excess work done on the entire chain (calculated as the difference between average energies at the end and at the beginning), S represents the excess work done only on the manipulated spin (calculated as the integral of the force applied on that spin) and APT represents the first-order adiabatic perturbation theory prediction (stripped of its oscillations). The main plot was constructed using the ground state of the total Hamiltonian~\eqref{eq:3spinHamiltonian} at $t_i$ as the initial state, thus leading to $W_\mathrm{ex} \sim \tau^{-2}$.
Meanwhile, the inset was constructed using a tensor product state: the ground state of the manipulated spin times the ground state of the other two spins combined, thus leading to $W_\mathrm{ex} \sim \tau^{-1}$.
}

\end{figure}

\subsection{Example: small Ising chain}

To illustrate our finding, we have included Fig.~\ref{fig:3spins}, which shows $W_\mathrm{ex}$ vs $\tau$ when we manipulate only one spin of an Ising chain with three spins, with Hamiltonian
\begin{equation} \label{eq:3spinHamiltonian}
H(\field_3) = - \frac{1}{2} \left( J \sum_{j=1}^{3} \sigma^z_j \sigma^z_{j+1} + \sum_{j=1}^{2} \field_j \sigma^x_j \right) - \frac{\field_3}{2} \sigma^x_3 .
\end{equation}
The magnetic fields $\field_1$ and $\field_2$ are constant, while the magnetic field $\field_3(t)$ of the manipulated spin is varied linearly in time.
For the main plot, we chose the ground state of Eq.~\eqref{eq:3spinHamiltonian} as the initial state.
Hence the excess work on the entire chain decays as $\tau^{-2}$ for large $J\tau$.
The excess work on the manipulated spin behaves exactly the same (apart from numerical errors).
On the other hand, in the inset, the initial state was chosen to be the tensor product of the ground state of spin $3$ and the ground state of spins $1$ and $2$ together, which is still a pure state but noncomutable with the initial Hamiltonian.
In this case, the excess work decays as $\tau^{-1}$ for large $J\tau$, which agrees with the APT prediction for nondiagonal initial states.
The notable oscillations are natural in such a small chain and they are predicted by APT, but we removed them from the APT result in this figure for ease of representation (see Ref.~\cite{Soriani2022PRA}).

\subsection{Example: driven quantum Brownian motion \label{sec:OpenSysCL}}
As a second illustrative example, we consider quantum Brownian motion in a harmonic trap whose minimum location varies in time. The Hamiltonian we use to model this phenomenon is given by
\begin{equation}
    H_{\mathrm{BM}}(\lambda(t)) = \frac{P^{2}}{2 M} + \kappa_{0}\frac{(Q-\lambda(t))^{2}}{2} + H_{\mathrm{CL}}\,,
    \label{eq:HamiltonianDBM}
\end{equation}
where $P$, $Q$ and $M$ denote, respectively, the momentum and position operators of the Brownian particle and its mass whereas $\kappa_{0}$ denotes the stiffness of the trap. By $\lambda(t)$, we denote the time-varying position of the minimum and $H_{\mathrm{CL}}$ stands for the Caldeira-Leggett model \cite{caldeiraBook},
\begin{equation}
    H_{\mathrm{CL}} = \sum_{k=1}^{N} \left[ \frac{p_{k}^{2}}{2 m_{k}} + \frac{m_{k} \omega_{k}^{2}}{2}(q_{k} - Q)^{2}\right]\,,
    \label{eq:HamiltonianCL}
\end{equation}
which describes a collection of $N$ harmonic oscillators with mass $m_{k}$, frequency $\omega_{k}$ and position and momentum operators $q_{k}$ and $p_{k}$ interacting with the Brownian particle.

According to Eq.~\eqref{eq:OpenSystemWork}, the work performed reads
\begin{eqnarray}
    W^{\mathrm{BM}} &=& \int_{t_{i}}^{t_{f}}dt\,\dot{\lambda}(t) \mathrm{Tr}\left\{\rho(t)  \frac{\partial H_{\mathrm{BM}}}{\partial\lambda}\right\}\nonumber\\
    &=& -\kappa_{0}\int_{t_{i}}^{t_{f}}dt\,\dot{\lambda}(t)\left(\langle Q(t)\rangle-\lambda(t)\right)\,,
    \label{eq:WorkBM}
\end{eqnarray}
where the non-equilibrium average $\langle Q(t)\rangle$,
\begin{equation}
    \langle Q(t) \rangle = \frac{\kappa_{0}}{M}\int_{t_{i}}^{t}dt'\,\Phi_{BM}(t-t') \lambda(t')\,,
    \label{eq:NoneqQCL}
\end{equation}
(see App.~\ref{sec:CL}) was obtained using the equilibrium initial state
\begin{equation}
    \rho(t_{i}) = Z_{i}^{-1}\exp{\left( -\beta H_{\mathrm{BM}}(\lambda(t_{i}))\right)}\,,
\end{equation}
where $Z_{i} = \mathrm{Tr}[\exp{(-\beta H_{\mathrm{BM}}(\lambda(t_{i})))}]$ and $\beta$ is the inverse temperature.

Inserting expression \eqref{eq:NoneqQCL} in Eq.~\eqref{eq:WorkBM} and performing an integration by parts, it is straightforward to show that
\begin{equation}
    W_{\mathrm{ex}}^{\mathrm{BM}} = \frac{\kappa_{0}^{2}}{M}\int_{t_{i}}^{t_{f}}\int_{t_{i}}^{t}\dot{\lambda}(t)\dot{\lambda}(t') \Psi_{\mathrm{BM}}(t-t')\,dt dt'\,,
    \label{eq:ExcessWorkBM}
\end{equation}
where $\Phi_{\mathrm{BM}}(t) = -d\Psi_{\mathrm{BM}}dt$. Hence, the excess work is exactly given by the LRT functional \eqref{eq:LRTExcessWork} derived in Sec.~\ref{sec:WeakProcesses}.

To investigate the large $\tau$ behavior of Eq.\eqref{eq:ExcessWorkBM}, one can first take the limit $N\to\infty$. In this case, an explicit expression for the so-called spectral density $J(\omega)$ has to be chosen in the continuum limit. Using the standard gapless Ohmic spectral density, 
\begin{equation}
   J(\omega) = \eta\omega\,,
\end{equation}
with a high-frequency cutoff, one obtains an exponentially decaying $\Psi_{\mathrm{BM}}(t)$ (see App.~\ref{sec:CL} for the details),
\begin{equation}
    \Psi_{\mathrm{BM}}(t) = \frac{e^{-\gamma t}}{\omega_{0}^{2}}\left[ \cos{(\Omega t)} + (\gamma/\Omega)\sin{(\Omega t)}\right]\,,
    \label{eq:RelaxFuncBM}
\end{equation}
where $\omega_{0}^{2}=\kappa_{0}/M$, $\Omega^{2} = \omega_{0}^{2}-\gamma^{2}$ and $\gamma = \eta/M$.

Plugging Eq.~\eqref{eq:RelaxFuncBM} in Eq.~\eqref{eq:ExcessWorkBM}, the excess work $W_{\mathrm{ex}}^{\mathrm{BM}}$ for the linear protocol, $\lambda(t) = \lambda_{0}+\delta\lambda\,(t/\tau)$, reads,
\begin{eqnarray}
    \lefteqn{W_{\mathrm{ex}}^{\mathrm{BM}} = \kappa_{0} (\delta\lambda)^{2} \left( \frac{\Omega}{\omega_{0}}\right)^{2}}\nonumber\\
    &\times&\left\{ 2\left(\frac{\gamma}{\omega_{0}} \right) \frac{[(\gamma/\Omega)+1]}{\omega_{0}\tau} + \frac{[1-3(\gamma/\Omega)^{2}]}{(\omega_{0}\tau)^{2}} + O(e^{-\gamma\tau})\right\}\,, \nonumber\\
\end{eqnarray}
where the leading order is $\tau^{-1}$.

On the other hand, if $N$ is large but kept finite, $\Psi_{\mathrm{BM}}(t)$ will be a quasi-periodic function as expressed in Eq.~\eqref{eq:RelaxationFunction}. In this case, the excess work \eqref{eq:ExcessWorkBM} for the same linear protocol will be given by a sum of terms proportional to
\begin{equation}
   \frac{1}{\tau^{2}} \int_{t_{i}}^{t_{f}}\int_{t_{i}}^{t}\cos{[\omega_{m n}(t-t')]}dt dt' = \frac{[1-\cos(\omega_{m n}\tau)]}{(\omega_{m n}\tau)^{2}}\,,
\end{equation}
where $\omega_{m n} = E_{m n}/\hbar$ are the frequencies related to the spectral gaps of the isolated system, as defined before in Sec.~\ref{sec:ExcessWork}. The previous expression shows that the $\tau^{-1}$ term is absent and suggests that the limits of large $\tau$ and large $N$ might not commute. This is particularly relevant for numerical simulations where $N$ is always finite.

\section{Discussion \label{sec:Discussion}}

In this section, we summarize the assumptions we made and the assumptions \emph{we did not make} to arrive at the asymptotic behavior of the excess work for large process duration, namely, $W_\mathrm{ex} \sim \tau^{-2}$ for a generic protocol.

It is clear from Sec.~\ref{sec:SlowProcesses} that the important ingredients are Hamiltonian dynamics and isolated gapped systems at temperatures well below their gap or open systems treated as part of a larger closed gapped system (where APT can be applied). Furthermore, we also assume initially diagonal density matrices.
This second condition reflects initial thermodynamic equilibrium $\dot\rho(t_i) = 0$, as per von Neumann's equation.
Note that tensor products of diagonal states of open system and environment \emph{do not} lead to diagonal states of the total composite system, given the presence of interactions between the parts.
Although such product states are often cited as equilibrium states when the open system is weakly coupled to the environment (as in the context of the geometric approach~\cite{Sivak2012,Bonanca2014,Zulkowski2015,Scandi2019}), the mild effects of the interaction can accumulate over long process durations and thus should not be ignored.
Whenever the initial state is not diagonal, the excess work does scale asymptotically as $\tau^{-1}$, as seen in the inset of Fig.~\ref{fig:3spins}.

As mentioned previously, our arguments are independent of the size of the system.
Moreover, we assumed no relaxation mechanism to be present, such as observed in systems obeying the eigenstate thermalization hypothesis or in the chaotic regime \cite{RigolAPH2016,BorgonoviPRP2016}.
This means our approach describes the asymptotic behavior of the excess work in any gapped system initially in thermodynamic equilibrium.
Of course, that does not rule out the possibility of other excess work scalings happening before the asymptotic behavior takes place --- indeed, two of us verified this in Ref.~\cite{Soriani2022PRA}.
In any case, with or without relaxation, the excess work eventually behaves as $\tau^{-2}$ for $\tau$ large enough in gapped systems.   

A connection between our analysis and relaxation mechanisms can be made through Eq.~\eqref{eq:LRTExcessWork2} written in the following form
\begin{equation}
    W_{\mathrm{ex}}^{\mathrm{LRT}} = \frac{\Delta^{2}}{2}\int_{0}^{1}\int_{0}^{1}\dot{g}(s)\dot{g}(s') \Psi_{i}\left(\tau(s-s')\right) ds ds'\,,
    \label{eq:LRTExcessWork3}
\end{equation}
where the integration variables were changed to $s=t/\tau$ and $s'=t'/\tau$, and the derivatives are now $\dot{g}(s)=dg/ds$ and $\dot{g}(s')=dg/ds'$.
It is important to stress that $\Psi_{i}(t)$ can either describe few-body or many-body correlations depending on the observable that couples to the control parameter.
Equation~\eqref{eq:LRTExcessWork3} suggests that the excess work scaling for weak and slowly-varying processes (the large $\tau$ limit) is ruled by the long time tail of $\Psi_{i}(t)$ (or, equivalently, by the small frequency behavior of its spectrum), which in turn is related to the long time tail of the response function (see Sec.~\ref{sec:WeakProcesses}).
It was verified in Sec.~\ref{sec:OpenSysCL} that, using Eq.~\eqref{eq:LRTExcessWork3} for linear protocol, a simple exponential decay leads to an asymptotic $\tau^{-1}$ scaling.
This type of response decay can be obtained in general from Lindblad master equations.
However, we should keep in mind that large times predictions of this effective description eventually fail (see Ref.~\cite{KlattPRL2021} for a recent discussion).

Although the TI chain, the system we used to corroborate our findings, is integrable and consequently nonrelaxing, the metric of the geometric approach can still be suitably defined for it.
For example, following the original LRT derivation of the geometric approach, the metric in the paramagnetic phase of the chain can be written as~\cite{Sivak2012,Bonanca2014}
\begin{equation} \label{eq:MetricFromLRT}
\zeta(B) = \Psi(B;0) \tau_{R}(B),
\end{equation}
where $\Psi(B;t)$ is the relaxation function of Eq.~\eqref{eq:ParamagneticRelaxFunction} with $B_i \to B$, $\Psi(B,0) = N J^2 / 4 B^3$ and the so called relaxation time is
\begin{equation} \label{eq:RelaxationTime}
\tau_{R}(B) = \int_{0}^{\infty} \frac{\Psi(B;t)}{\Psi(B;0)} dt.
\end{equation}
This integral vanishes when the full paramagnetic relaxation function \eqref{eq:ParamagneticRelaxFunction} is used, owning to the presence of the oscillatory cosine factor.
Heuristically, we might argue that these oscillations should not contribute to a measure of relaxation of the system, and hence we can drop the cosine factor in this calculation.
Considering just the Bessel envelope of Fig.~\ref{fig:LRT_RelaxationFunction}, we arrive at $\tau_{R} = J^{-1}$, which is the natural decay time scale of the Bessel factor.
Nevertheless, simply putting Eq.~\eqref{eq:MetricFromLRT} into Eq.~\eqref{eq:GeoExcessWork} does not reproduce the correct excess work in any of the numerical simulations shown in this paper.

\section{Conclusion \label{sec:Conclusion}}

Despite increased recent interest in the development of a unifying theoretical framework to investigate the minimization of energetic costs in finite-time thermodynamic processes of quantum systems \cite{abiuso2020entropy}, it seems that such unification still evades us.
Here, we have provided extensive evidence to support the claim that the celebrated geometric approach to optimal driving, in its current formulation, does not apply to quantum gapped systems evolving under unitary dynamics.
As far as adiabatic perturbation theory and linear response theory reach, there is no sign of a thermodynamic metric from which geodesics can be obtained, neither is there the predicted $\tau^{-1}$ asymptotic behavior of the excess work.
However, our analysis does not exclude the possibility of studying the so-called thermodynamic length along the lines of Refs.~\cite{deffner2010prl,deffner2013pre}.
Attempts at a geometric approach for isolated quantum systems can be found in Refs.~\cite{Tomka2016,Chen2022}.

Although some elements of our numerical analysis of the transverse-field Ising chain may be system-specific, we expect most prominent features highlighted in this paper to be universal.
In the context of slowly-varying processes, we have discussed and exemplified the minimum exponent decay $\tau^{-2}$ of the excess work, as guaranteed by adiabatic perturbation theory.
We then tested two well-known optimization strategies, enabling us to indefinitely increase the exponent or ensure the minimum exponent decay for reasonably small process duration $\tau$.
In the complementary regime of fast but weak processes, we applied a Chebyshev expansion approach to optimization, which is based only on the knowledge of equilibrium correlation functions of the system, in typical linear response theory fashion.
The protocols obtained are surprisingly sensitive to the time scales present in the relaxation function, and they reproduce the boundary cancellation method in the slow limit.

In both regimes studied, the perturbation theories do not give lower bounds for the excess work of the protocols generated: higher order BCM protocols can always be used to guarantee steeper decay in the APT regime, just as higher order Chebyshev expansions seem to guarantee ever-increasing performance in the LRT regime.
Overall, the present results seem to suggest that, in isolated quantum systems, there is no single way of obtaining optimal protocols for every scenario and, as a matter of fact, it is not clear if the notion of a unique optimization procedure exists.

Finally, we have briefly analyzed the consequences of our findings to open quantum systems with two simple but illustrative examples, showing that the asymptotic $\tau^{-1}$ scaling is indeed absent for initially diagonal states and that the limits of large $N$ and large $\tau$ might not commute.
More detailed work is however necessary to understand how the geometric approach predictions can be reconciled with the results presented here.
This further analysis might focus on the role of the initial state of the full isolated system and on the reliability of the long time predictions of effective descriptions such as Lindblad master equations.
With little effort to go beyond what we have shown here for initially diagonal states, it can be shown that an initial state that does not commute with the full initial Hamiltonian leads to a non-zero first-order APT correction to the excess work (as demonstrated in the inset of Fig.~\ref{fig:3spins}).
Concerning Lindblad master equations, they often imply exponential decays of correlation functions which, according to our LRT approach, would lead to an asymptotic $\tau^{-1}$ scaling in the large $\tau$ limit.

Our analyses certainly leave several important open questions concerning the conditions required for the $\tau^{-1}$ scaling of the excess work.
In this sense, this paper consists of a critical assessment of the claims that have been made so far within the geometric approach in quantum systems.
In particular, predictions coming from effective descriptions of open quantum systems often employ phenomenological assumptions that should be very critically analyzed, especially those that can impact the scaling discussed here.
It remains to be proven then under which conditions truly Hamiltonian quantum dynamics might lead to the $\tau^{-1}$ scaling.

\begin{acknowledgments}
The authors acknowledge M. Perarnau-Llobet for fruitful discussions on the topic.
A.S. and M.V.S.B. thank the National Council for Scientific and Technological Development (CNPq, Brazil) under Grant No. 140549/2018-8 and FAEPEX (Fundo de Apoio ao Ensino, \`a Pesquisa e \`a Extens\~ao, Brasil), Grant No. 2146-22.
M.V.S.B. acknowledges financial support by FAPESP (Funda\c{c}\~ao de Amparo \`a Pesquisa do Estado de S\~ao Paulo)(Grant No. 2020/02170-4).
E.M. also thanks the support of CNPq through Grant No. 309584/2021-3 and Capes through Grant No. 0899/2018.
\end{acknowledgments}

\appendix

\section{Transverse-field Ising chain \label{sec:TIchain}}

The transverse field Ising model (TI) \cite{Pfeuty1970} is a one-dimensional chain with $N$ spins and first-neighbor interactions.
Its Hamiltonian is
\begin{equation} \label{eq:TIHamiltonian}
H_{\mathrm{TI}}(\field) = - \frac{1}{2} \left( J \sum_{j=1}^{N} \sigma^z_j \sigma^z_{j+1} + \field \sum_{j=1}^N \sigma^x_j \right),
\end{equation}
where $\sigma_j^{x,z}$ are standard Pauli matrices for each spin $j$ (with $\sigma_{N+1}^{x,z} = \sigma_{1}^{x,z}$), $J$ is the coupling constant and $\field$ is the external magnetic field (the $1/2$ global factor was added for later convenience).
We work in units such that $\hbar=1$.
For simplicity, we assume $N$ to be even and that the system is initially in its ground state.
After a Jordan-Wigner transform, a Fourier transform and Bogoliubov transform \cite{Pfeuty1970}, Eq.~\eqref{eq:TIHamiltonian} is brought to diagonal form, represented by non-interacting fermions with dispersion
\begin{equation} \label{eq:TIDispersion}
\epsilon_k(\field) = \sqrt{\left( \field - J\cos k \right)^2 + J^2 \sin^2 k},
\end{equation}
for $N$ allowed values of momentum $k = (2n + 1) \pi/N$, given integer $n$ between $-N/2$ and $N/2 - 1$.
In the thermodynamic limit, $k$ is a continuous variable ranging from $-\pi$ to $\pi$, and sums over $k$ are replaced by integrals.

The dynamics of the system, when initially prepared in the ground state, can be simplified into the dynamics of $N/2$ two-level systems (known as Landau-Zener systems), one for each positive value of $k$ \cite{Dziarmaga2005}.
The evolved ground state can be written as
\begin{equation} \label{eq:TIEvolvedGroundState}
\ket{\psi(t)} = \bigotimes_{k>0} \Bigl( u_k(t) \ket{\downarrow_k} - v_k(t) \ket{\uparrow_k} \Bigr),
\end{equation}
where $\ket{\uparrow_k}$ and $\ket{\downarrow_k}$ form a basis of the two-level system labeled by $k$.
Placing Eq.~\eqref{eq:TIEvolvedGroundState} into Schrödinger's equation leads to (omitting time-dependencies)
\begin{equation} \label{eq:TIDifferentialEquations}
\begin{split}
i\, \dot{u}_k & =  - \left( \field - J \cos k \right) u_k - J \sin k \, v_k, \\
i\, \dot{v}_k & =  - J \sin k \, u_k + \left( \field - J \cos k \right) v_k.
\end{split}
\end{equation}
The numerical results presented in this manuscript were obtained from the standard fourth-order Runge-Kutta method applied to Eqs.~\eqref{eq:TIDifferentialEquations}.

The excess work per spin is obtained from Eq.~\eqref{eq:ExcessWork},
\begin{equation} \label{eq:TIwork}
w_{\mathrm{ex}}(\tau) = \frac{1}{N} \sum_{k>0} 2\epsilon_k \left| u_k \sin\frac{\theta_k}{2} - v_k \cos\frac{\theta_k}{2} \right|^2,
\end{equation}
where all quantities should be evaluated at $t_f$, including
\begin{equation} \label{eq:TItheta}
\theta_k(\field) = \arctan\left( \frac{J \sin k}{\field - J \cos k} \right).
\end{equation}
The system's relaxation function is obtained straight from Eq.~\eqref{eq:RelaxationFunction},
\begin{equation} \label{eq:TIRelaxFunction}
\Psi(\field;t) = J^2 \sum_{k>0} \frac{\sin^2 k}{\epsilon_k^3(\field)} \cos\bigl( 2\epsilon_k(\field) t \bigr).
\end{equation}

\section{Agreement between LRT and APT \label{sec:APTandLRTagree}}

Our goal in this appendix is to show that, for processes that are both weak and slow, the expressions for the excess work from LRT and from APT are identical.

To begin, consider first Eq.~\eqref{eq:LRTExcessWork2} for the excess work calculated by means of LRT --- it already expresses the leading order of the excess work in $\Delta$ when $\Delta \to 0$.
We want to find its adiabatic limit, that is, the leading order in $\tau^{-1}$ for $\tau \to \infty$.
To this end, we can perform an integration by parts in the inner integral of Eq.~\eqref{eq:LRTExcessWork2},
\begin{multline*}
\int_{t_i}^{t_f} \dot{g}(t') \Psi_i(t-t') dt' = \dot g(t_f) \int_{t}^{t_f} \Psi_i(t-t'') dt''
\\
- \dot g(t_i) \int_{t}^{t_i} \Psi(t-t'') dt'' - \int_{t_i}^{t_f} \ddot g(t') \int_{t}^{t'} \Psi_i(t-t'') dt'' dt'
\end{multline*}
and note that the last term on the right-hand side is of a higher order in $\tau^{-1}$ than the first two, since $\dot g(t) \propto \tau^{-1}$ and $\ddot g(t) \propto \tau^{-2}$.
Hence, as long as the relaxation function $\Psi_i$ is composed of sums of oscillatory functions [as in Eqs.~\eqref{eq:RelaxationFunction}, \eqref{eq:ParamagneticRelaxFunction} and \eqref{eq:TIRelaxFunction}], we can drop the last term, and the same applies to the outer integral in Eq.~\eqref{eq:LRTExcessWork2}.
If this is done consistently, the LRT excess work reduces to
\begin{multline} \label{eq:LRTExcessWork_Adiabatic}
W_{\mathrm{ex}}^\mathrm{LRT}(\tau) \approx \frac{\Delta^2}{2} \bigg( \left( \dot{g}^2(t_f) + \dot{g}^2(t_i) \right)  \Upsilon_i(0)\\
- 2 \dot{g}(t_f) \dot{g}(t_i) \Upsilon_i(\tau) \bigg),
\end{multline}
where we introduced the function $\Upsilon$, whose relation to $\Psi$ is given by $\Psi(\lambda;t) = - \pd_t^2 \Upsilon(\lambda;t)$.
From Eq.~\eqref{eq:RelaxationFunction} we get
\begin{equation} \label{eq:LRTUpsilon}
\Upsilon(\lambda;t) = 2\hbar^2 \sideset{}{'}\sum_{m,n} p_n \frac{|F_{mn}(\lambda)|^2}{E_{mn}^3(\lambda)} \cos\left( \frac{E_{mn}(\lambda)}{\hbar} t \right)
\end{equation}
and $\Upsilon_i(t) = \Upsilon(\lambda_i;t)$.

On the other hand, consider Eq.~\eqref{eq:APT_W2} for the excess work calculated by means of APT --- it gives the leading order of the excess work in $\tau^{-1}$ when $\tau \to \infty$.
Similar to above, we want to find the weak limit of this equation, that is, its leading order in $\Delta$ for $\Delta \to 0$.
In order to do so, we first place Eqs.~\eqref{eq:APT_FirstOrderCoefficient} and \eqref{eq:APT_Matrix} into Eq.~\eqref{eq:APT_W2},
\begin{multline*}
W^\mathrm{APT}(\tau) = \hbar^2 \sideset{}{'}\sum_{m,n} p_n E_{mn}(\lambda_f) \\
\times \left| \frac{F_{mn}(\lambda_f)}{E_{mn}^2(\lambda_f)} \dot\lambda(t_f) - e^{i \phi_{mn}(t_f)} \frac{F_{mn}(\lambda_i)}{E_{mn}^2(\lambda_i)} \dot\lambda(t_i) \right|^2.
\end{multline*}
Now, since $\Delta$ is small, $\lambda_f \approx \lambda_i$ and we can replace every instance of the former with the latter (including in the energy gap appearing in the adiabatic phases, inside the integral of Eq.~\eqref{eq:AdiabaticPhases}).
Doing this carefully leads to (with $\dot \lambda = \Delta \dot g$)
\begin{multline} \label{eq:APTExcessWork_Weak}
W_\mathrm{ex}^\mathrm{APT}(\tau) \approx \hbar^2 \Delta^2 \sideset{}{'}\sum_{m,n} p_n \frac{ | F_{mn}(\lambda_i) |^2 }{E_{mn}^3(\lambda_i)} \\
\times \left| \dot{g}(t_f)  - e^{E_{mn}(\lambda_i) \tau/i\hbar} \dot{g}(t_i) \right|^2,
\end{multline}
Lastly, expanding the squared absolute value in Eq.~\eqref{eq:APTExcessWork_Weak} reveals that this expression is indeed equal to Eq.~\eqref{eq:LRTExcessWork_Adiabatic}.

Both LRT and APT predict an excess work that is quadratic in their small perturbative parameter ($\Delta$ and $\tau^{-1}$, respectively).
Their agreement in the simultaneously weak and slow regime --- which can be shown to hold for the microscopic state of the system, instead of the excess work --- is a consequence of the fact that both are well defined perturbation theories for the same equation, namely, Schrödinger's equation.

\section{Driven quantum Brownian motion \label{sec:CL}}

In this appendix we sketch the main steps to obtain Eqs.~\eqref{eq:NoneqQCL} and \eqref{eq:RelaxFuncBM}. We start from Eqs.~\eqref{eq:HamiltonianDBM} and \eqref{eq:HamiltonianCL} and the corresponding Heisenberg equations to obtain
\begin{equation}
    \ddot{q}_{k} + \omega_{k}^{2} q_{k} = \omega_{k}^{2}Q\,,
    \label{eq:SoluQk}
\end{equation}
and
\begin{multline} \label{eq:EquationQ1}
\ddot{Q} + \omega_{0}^{2} Q = \\
\omega_{0}^{2}\lambda(t) - \frac{1}{M}\left( \sum_{k=1}^{N}m_{k}\omega_{k}^{2}\right) Q + \frac{1}{M}\sum_{k=1}^{N}m_{k}\omega_{k}^{2} q_{k},
\end{multline}
where $\omega_{0}^{2}=\kappa_{0}/M$.

Plugging the solution of Eq.~\eqref{eq:SoluQk},
\begin{multline}
q_{k}(t) = q_{k}(0)\cos{(\omega_{k} t)} + \frac{p_{k}}{m_{k}\omega_{k}}\sin{(\omega_{k} t)} \\
    + \omega_{k} \int_{0}^{t}dt'\sin{[\omega_{k}(t-t')]} Q(t')\,,
\end{multline}
into Eq.~\eqref{eq:EquationQ1}, the equation of motion for $Q$ reads
\begin{equation} \label{eq:EquationQ2}
\ddot{Q} + \omega_{0}^{2} Q + \frac{1}{M}\int_{0}^{t}dt' \chi(t-t') \dot{Q}(t') =\omega_{0}^{2}\lambda(t) + \frac{f_{1}(t)}{M}
\end{equation}
where we have defined
\begin{equation}
    \chi(t) = \sum_{k=1}^{N} m_{k}\omega_{k}^{2} \cos{(\omega_{k}t)}\,,
    \label{eq:DefinitionChi}
\end{equation}
and
\begin{multline}
f_{1}(t) = \sum_{k=1}^{N}\Bigl[ m_{k}\omega_{k}^{2} \bigl( q_{k}(0)-Q(0) \bigr) \cos{(\omega_{k} t)} \\
+ \omega_{k} p_{k}\sin{(\omega_{k}t)}\Bigr]\,.
\end{multline}

The solution of Eq.~\eqref{eq:EquationQ2} can be expressed then as
\begin{multline} \label{eq:SolutionFQ}
Q(t) = \left( \frac{d\Phi_{\mathrm{BM}}(t)}{dt}+\Phi_{\mathrm{BM}}(0)\right) Q(0) +\Phi_{\mathrm{BM}}(t)\frac{P(0)}{M} \\
+ \frac{1}{M}\int_{0}^{t} dt' \Phi_{\mathrm{BM}}(t-t') f_{2}(t') \\
+ \omega_{0}^{2}\int_{0}^{t} dt' \Phi_{\mathrm{BM}}(t-t') \lambda(t')\,,
\end{multline}
where
\begin{equation}
    f_{2}(t) = f_{1}(t) + Q(0) \chi(t)
\end{equation}
and
\begin{equation}
    \Phi_{\mathrm{BM}}(t) = \mathcal{L}^{-1} \left\{ \left[ s^{2} + s(\tilde{\chi}(s)/M)+\omega_{0}^{2}\right]^{-1}\right\}\,,
    \label{eq:RespFuncBM}
\end{equation}
with $\mathcal{L}^{-1}\{F(s)\}$ denoting the inverse Laplace transform of $F(s)$.
By $\tilde{\chi}(s)$, we denote the Laplace transform of $\chi(t)$ given by Eq.~\eqref{eq:DefinitionChi}.

Taking the trace of Eq.~\eqref{eq:SolutionFQ} over the initial equilibrium state,
\begin{equation}
    \rho(0) = Z^{-1}\exp{\left(-\beta H_{\mathrm{BM}}(0) \right)}\,,
\end{equation}
with $Z = \mathrm{Tr}\exp{(-\beta H_{\mathrm{BM}}(0))}$, $\beta = (k_{B}T)^{-1}$ the inverse temperature and $H_{\mathrm{BM}}(0)$ given by Eq.~\eqref{eq:HamiltonianDBM} with $\lambda(0)=0$, we finally obtain the non-equilibrium average value,
\begin{equation}
    \langle Q(t) \rangle = \omega_{0}^{2}\int_{0}^{t}dt' \Phi_{\mathrm{BM}}(t-t') \lambda(t')\,.
\end{equation}

If the limit $N\to\infty$ is taken, it is convenient to introduce a continuous expression for the so-called spectral density $J(\omega)$ \cite{caldeiraBook},
\begin{equation}
    J(\omega)= \frac{\pi}{2}\sum_{k=1}^{N} m_{k}\omega_{k}^{3} \delta(\omega-\omega_{k})\,,
\end{equation}
where $\delta(.)$ denotes Dirac's delta function.
In the standard case of an Ohmic spectral density, $J(\omega)$ reads \cite{caldeiraBook}
\begin{equation}
    J(\omega) = \eta \omega\,,\; \mathrm{if}\, \omega \leq \omega_{D}\,,
    \label{eq:SpectralDOhmic}
\end{equation} 
where $\omega_{D}$ is a high-frequency cutoff.
In terms of $J(\omega)$, the function $\chi(t)$ is given by
\begin{equation}
    \chi(t) = \frac{2}{\pi}\int_{0}^{\infty}d\omega \frac{J(\omega)}{\omega} \cos{(\omega t)} = 2\eta \delta(t)\,,
\end{equation}
where the last equality was obtained using Eq.~\eqref{eq:SpectralDOhmic} and taking the limit $\omega_{D}\to\infty$.

The previous expression for $\chi(t)$ implies $\tilde{\chi}(s)=2\eta$.
Inserting this expression in Eq.~\eqref{eq:RespFuncBM} and defining $\gamma=\eta/M$, we obtain
\begin{equation}
    \Phi_{\mathrm{BM}}(t) = \frac{e^{-\gamma t}\sin{(\Omega t)}}{\Omega}\,,
\end{equation}
where $\Omega^{2} = \omega_{0}^{2} - \gamma^{2}$.
From this expression for $\Phi_{\mathrm{BM}}(t)$, we obtain $\Psi_{\mathrm{BM}}(t)$ in \eqref{eq:RelaxFuncBM} knowing that $\Phi_{\mathrm{BM}}(t)=-d\Psi_{\mathrm{BM}}(t)/dt$.

\bibliography{bibliography}

\end{document}